\documentclass[journal]{new-aiaa}

\usepackage[utf8]{inputenc}
\usepackage{textcomp}
\usepackage{graphicx}
\usepackage{amsmath}
\usepackage[version=4]{mhchem}
\usepackage{siunitx}
\usepackage{longtable,tabularx}
\usepackage{booktabs} 
\usepackage{subfig}
\usepackage{subfloat}
\usepackage{epstopdf}

\newcommand{\norm}[1]{\left\Vert#1\right\Vert} 
\newcommand{\bbm}{\begin{bmatrix}}
\newcommand{\ebm}{\end{bmatrix}}
\newcommand{\vect}{\underrightarrow}
\DeclareMathAlphabet{\mbf}{OT1}{ptm}{b}{n}
\newcommand{\mbs}[1]{{\boldsymbol{#1}}}
\newcommand{\trans}{{\ensuremath{\mathsf{T}}}} 
\newcommand{\beq}{\begin{equation}}
\newcommand{\eeq}{\end{equation}}
\newcommand{\beqarray}{\begin{eqnarray}}
\newcommand{\eeqarray}{\end{eqnarray}}

\title{Atmospheric Density-Compensating Model Predictive Control for Targeted Reentry of Drag-Modulated Spacecraft}

\author{Alex D. Hayes\footnote{Ph.D. Student, Department of Aerospace Engineering \& Mechanics, 110 Union St. SE, Minneapolis, MN 55455, email: hayes455@umn.edu.} and Ryan J. Caverly\footnote{Assistant Professor, Department of Aerospace Engineering \& Mechanics, 110 Union St. SE, Minneapolis, MN 55455, AIAA Member, email: rcaverly@umn.edu (Corresponding Author)}}
\affil{University of Minnesota, Twin Cities, Minneapolis, MN 55455}

\begin{document}

\maketitle

\begin{abstract}
This paper presents an estimation and control framework that enables the targeted reentry of a drag-modulated spacecraft in the presence of atmospheric density uncertainty. In particular, an extended Kalman filter is used to estimate errors between the in-flight atmospheric density density and the atmospheric density used to generate the guidance trajectory. This information is leveraged within a model predictive control strategy to improve tracking performance, reduce control effort, and increase robustness to actuator saturation compared to the state-of-the-art approach. The estimation and control framework is tested in a Monte Carlo simulation campaign with historical space weather data. These simulation efforts demonstrate that the proposed framework is able to stay within 100 km of the guidance trajectory at all points in time for 98.4\% of cases. The remaining 1.6\% of cases were pushed away from the guidance by large density errors, many due to significant solar storms and flares, that could not physically be compensated for by the drag control device. For the successful cases, the proposed framework was able to guide the deorbiting spacecraft to the desired location at the entry interface altitude with a mean error of 12.1 km and 99.7\% of cases below 100 km.
\end{abstract}

\section{Introduction}

\lettrine{T}{he} use of small spacecraft such as CubeSats~\cite{Chin2008} has become increasingly common in recent years, enabled by the miniaturization of electronics. These small spacecraft have brought with them a decrease in the cost of building and launching a satellite into orbit. The relatively low cost of these spacecraft has increased accessibility to space for educational institutions and opened up new business cases in low Earth orbit. However, due to the tight mass and volume constraints inherent in building a small spacecraft, it is difficult to incorporate a propulsion system in their design. 

Miniaturized propulsion systems that have been developed are expensive, which negates one of the primary benefits of building a small spacecraft - its low cost. Small spacecraft typically fly as secondary payload to reduce delivery costs. As a result, the spacecraft must undergo extensive testing to ensure that it does not pose a significant risk to the primary payload, which drives up its overall cost. The performance of small propulsion systems is also inadequate for making significant changes to the orbit of a spacecraft \cite{Pascoa2018}. For instance, as a result of these drawbacks, CubeSat designs rarely incorporate a propulsion system, resulting in a lack of maneuverability that reduces the applications for which very small spacecraft are suited.

While performing orbit changes using propulsion may not be practical for many small satellite missions, other methods have been proposed. A spacecraft operating in low-Earth orbit (LEO) will be flying through the upper portion of Earth's atmosphere in a free molecular flow regime due to the low density. As the spacecraft travels through the atmosphere, a small drag force is produced, which slowly decreases the orbital energy of the spacecraft, changing the orbit until the spacecraft eventually reenters the atmosphere~\cite{Vinh1980}. Researchers at NASA Ames Research Center are developing the ExoBrake, a device that enables a small spacecraft to modulate the drag force acting on the spacecraft by adjusting it's aerodynamic properties~\cite{Murbach2010}. A similar device is the Drag Deorbit Device being developed at the University of Florida~\cite{Guglielmo2019}. Modulating the drag force allows the effect of drag to be harnessed to provide some control over the trajectory of the spacecraft without the need for propellant. 

Drag modulation has been investigated to assist with maneuvering spacecraft in LEO. Examples of this include creating a differential drag between multiple satellites to fly in and maintain a formation~\cite{Leonard1989,kumar2011differential,ivanov2018study, Chocron2017} or perform constellation phasing maneuvers~\cite{bussy2018assessment,foster2018constellation,falcone2023propulsion}. Similarly, the rendezvous of satellites in orbit using drag modulation has also been studied~\cite{bevilacqua2008rendezvous, horsley2013small,harris2014minimum,dell2015optimal,mazal2016spacecraft}. Even if a spacecraft has propulsive capabilities, the use of the drag force produced during flight through an atmosphere may enable the spacecraft to use less propellant when changing orbits through an aeroassist~\cite{Vinh1986} or aerocapture maneuver~\cite{putnam2014drag,werner2019mission}. Another application of drag modulation, and the focus of this paper, is targeted reentry of spacecraft from low-Earth orbit, which can be used to safely dispose of satellites at the end of their service life and has been proposed as a method for returning small payloads to Earth~\cite{Murbach2010,lee2022development}. The HyCUBE concept~\cite{anderson2021preliminary,gardi2024simulation} is considering the use of drag modulation to deorbit a small, instrumented reentry vehicle in order to perform hypersonic aerothermodynamic testing. 

Several methods have been developed for generating nominal deorbit trajectories that target the entry interface location of a drag-modulated spacecraft~\cite{Virgili2015,Dutta2017,Omar2017conf,lee2022development,gaglio2023time}. These methods involve switching between a sequence of ballistic coefficients at discrete times in order to nominally reenter at a desired location relative to the surface of the Earth.
However, the drag force is dependant on the density of the local atmosphere, which is highly uncertain and variable~\cite{Doornbos2006}. This uncertainty in density and the corresponding drag force would cause the spacecraft to drift far away from the nominal trajectory if the nominal control policy were implemented in an open-loop fashion, likely resulting in large targeting errors at entry interface.

In order to compensate for atmospheric uncertainty, a new trajectory could be generated, however, the methods in~\cite{Virgili2015,Dutta2017,Omar2017conf,lee2022development,gaglio2023time} are computationally expensive and lack any guarantees of convergence, making them unsuitable for real-time implementation on-board the spacecraft. Additionally, if the spacecraft drifts from the nominal trajectory, a new trajectory to the desired terminal condition (e.g., a desired latitude and longitude at entry interface) may not exist. A better option is to use the nominal trajectory as the guidance for a closed-loop control strategy. In addition to developing an algorithm for generating the nominal drag-modulated reentry trajectory, Omar~\cite{Omar2019} utilized a linear quadratic regulator to track guidance. Model predictive control (MPC) has become a practical control method for spacecraft applications~\cite{eren2017model,di2018real,petersen2023safe}, largely motivated by its ability to explicitly handle state and input constraints. A typical MPC implementation involves solving a receding-horizon optimal control problem through online optimization. Although the implementation of MPC can require substantial computational resources, careful formulation of the optimization problem (e.g., solving a nonlinear MPC problem as a linear MPC problem with approximated dynamics) and the improvement in on-board spacecraft computational capabilities have helped make MPC a practical choice for spacecraft applications~\cite{di2018real}. This motivated the development of an MPC framework for drag-based deorbit tracking in~\cite{Hayes2022} that reduced tracking error and actuator usage by exploiting knowledge of the atmospheric density. Additionally, incremental MPC has been proposed to counter the effect of atmospheric density disturbances without the need for knowledge of the atmospheric density~\cite{la2024incremental}. 
Previous work~\cite{Hayes2022} demonstrated that MPC can provide improved tracking performance if knowledge of the atmospheric density was available but did not demonstrate a realistic method for obtaining such knowledge or provide a rigorous evaluation of the performance of the controller. And while the incremental MPC approach in~\cite{la2024incremental} can counter atmospheric disturbances without knowledge of the density, the incremental formulation requires that the model be discretized over small time interval in order to preserve the validity of the assumption that the disturbance is constant over the interval. As a result, the prediction horizon of the controller is limited by computational performance, which in turn limits the performance of MPC. These limitations demonstrate the need for an MPC approach that can incorporate atmospheric density estimates derived from realistic sensor measurements and can accommodate longer timesteps than those in~\cite{la2024incremental}, thus enabling longer prediction horizons and improved performance.

This paper presents a simple and efficient method for estimating the day-of-flight atmospheric density using an extended Kalman filter (EKF) with GPS measurements of the motion of the spacecraft relative to the guidance trajectory. Inspired by atmospheric density estimation methods for aerocapture~\cite{putnam2014drag,roelke2023atmospheric} and atmospheric entry~\cite{tracy2023robust,rea2024orion}, our proposed method estimates a scale factor on the nominal atmospheric density to compensate for day-of-flight density errors. This density estimation approach is combined with an improved version of the MPC formulation in~\cite{Hayes2022}, which is updated to include linear time-varying (LTV) system dynamics and an MPC objective function that penalizes use of the drag device in a more practical manner. The combined estimation and control framework is exhaustively tested in this paper through a Monte Carlo simulation campaign to provide a statistical measure of the performance of the proposed framework when subjected to a wide range of initial conditions, desired reentry locations, guidance trajectories and atmospheric density dispersions that are derived from historical observations. 

In summary, there are four main contributions in this paper that distinguish it from the work of~\cite{Hayes2022,la2024incremental}. The first is related to estimating the day-of-flight atmospheric density using an EKF with GPS measurements. By estimating the atmospheric density, atmospheric disturbances can be countered without the small discretization interval required by incremental MPC~\cite{la2024incremental}, enabling longer prediction horizons and unlocking the inherent performance benefits of MPC. The second contribution is the extension of the MPC formulation in~\cite{Hayes2022} to include LTV dynamics and a penalty on changing the drag device's ballistic coefficient, rather than penalizing the value of the ballistic coefficient. This extension results in a more accurate prediction model and a reduction in the amount of actuation needed from the vehicle's drag device. The third contribution is the thorough evaluation of the proposed estimation and control framework through Monte Carlo simulations using historical observations of space weather data to derive realistic atmospheric density variation. Such a simulation campaign demonstrates the effectiveness of the control framework when relying only on information that is realistically obtainable in a manner that is not considered in~\cite{Hayes2022,la2024incremental}. The fourth contribution of this paper is a slightly improved method to select the nominal ballistic coefficients in the longitude targeting portion of the guidance algorithms of~\cite{Omar2017conf,Omar2019}. Our proposed method helps center the guidance ballistic coefficients away from the saturation limits of the drag device, resulting in more control authority when tracking the guidance trajectory.

This paper is laid out as follows: first, preliminary information regarding reference frames and notation, atmospheric density and drag, and drag modulation is provided in Section~\ref{sec:Preliminaries}. Next, the methodology for generating guidance trajectories, estimating the atmospheric density and performing the model predictive control is discussed in Section~\ref{sec:Methodology}. Subsequently, the Monte Carlo simulation campaign is described and results are presented in Section~\ref{sec:Simulations}. Finally, conclusions and future work are given in Section~\ref{sec:Conclusion}.

\section{Preliminaries}
\label{sec:Preliminaries}

This section presents the definition of important reference frames, the concept of drag modulation and how it affects the acceleration of a spacecraft, as well as how atmospheric density error results in a change in drag acting on a spacecraft.

\subsection{Reference Frames}
\subsubsection{Earth-Centered Inertial}
As shown in Fig.~\ref{fig:LVLH}, the Earth-centered inertial ($ECI$) frame is defined by the basis vectors $\underrightarrow{ECI}^1$, $\underrightarrow{ECI}^2$ and $\underrightarrow{ECI}^3$. It is accompanied by the fixed point (unforced particle), $p$, which is defined at the center of the Earth. The nonlinear dynamics of the orbiting spacecraft are expressed in the $ECI$ frame, where the position and velocity of the spacecraft are found relative to point $p$, with time derivatives taken with respect to the $ECI$ frame. Nominal spacecraft trajectories are also described in this frame. The center of mass of the spacecraft is located at a point $s$ and the position of the spacecraft relative to the center of the Earth is expressed in the $ECI$ frame as $\mbf{r}^{sp}_{ECI}$, while the velocity of the spacecraft relative to the center of the Earth with respect to the $ECI$ frame is $\mbf{v}^{sp/ECI}_{ECI}$. Guidance trajectories are similarly expressed in this frame as $\mbf{r}^{gp}_{ECI}(t)$ and $\mbf{v}^{gp/ECI}_{ECI}(t)$, where point $g$ is the center of mass of a spacecraft on the guidance trajectory.
\subsubsection{Earth Centered, Earth Fixed}
The Earth centered, Earth fixed ($ECEF$) frame is defined by basis vectors that rotate along with the Earth with the $\vect{ECEF}^3$ direction pointing North, in the direction of the rotation of the Earth. This frame is used to locate spacecraft relative to the surface of the Earth in order to compute the local density of the atmosphere. Additionally, it is also assumed that the atmosphere is fixed in this frame, meaning winds are neglected. As a result, the velocity of the spacecraft expressed in the rotating $ECEF$ frame is the velocity of the spacecraft relative to the atmosphere. 

\subsubsection{Local Vertical, Local Horizontal}
\begin{figure}[t!]
	\centering
		\includegraphics[width=0.7\linewidth]{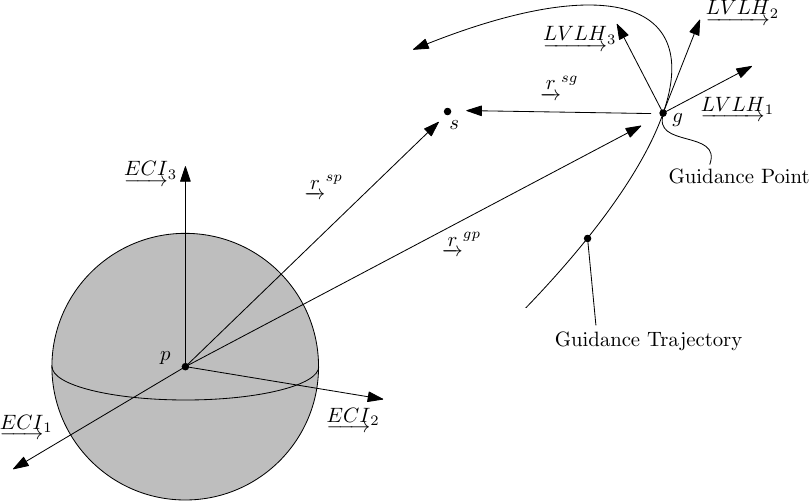} 
	\caption{The directions of the local vertical, local horizontal frame.}
	\label{fig:LVLH}
\end{figure}

The local vertical, local horizontal ($LVLH$) frame, shown in Fig.~\ref{fig:LVLH}, is used to describe the position and velocity of the spacecraft relative to a desired guidance position described by a point $g$.  This frame is also used to describe the dynamics of the spacecraft relative to the guidance trajectory. 
The directions of the $LVLH$ frame, expressed in the $ECI$ frame, are obtained using the guidance position and velocity, as expressed in the inertial frame. The $\underrightarrow{LVLH}^1$ direction points in the radial direction, from the center of the Earth to the guidance position. Expressed in the $ECI$ frame, this direction is obtained through 
	
\beq
\mbf{LVLH}^1_{ECI} = \frac{\mbf{r}^{~gp}_{ECI}}{\norm{\mbf{r}^{~gp}_{ECI}}_2}.
\eeq
The $\underrightarrow{LVLH}^3$ direction points in the direction of the angular momentum of the guidance trajectory. Expressed in the inertial frame, this direction is found through

\beq
\mbf{LVLH}^3_{ECI} = \frac{\left({\mbf{r}^{~gp}_{ECI}}\right)^\times \mbf{v}^{~gp/ECI}_{ECI}}{\norm{\left({\mbf{r}^{~gp}_{ECI}}\right)^\times \mbf{v}^{~gp/ECI}_{ECI}}_2},
\eeq 		
where $\left( \cdot \right)^\times$ is the skew-symmetric cross product operator~\cite{hughes2012spacecraft}. The $\underrightarrow{LVLH}^2$ direction completes a right-handed coordinate system through $\mbf{LVLH}^2_{ECI} = \left({\mbf{LVLH}^3_{ECI}}\right)^\times \mbf{LVLH}^1_{ECI}$
and points in the in-track direction of the guidance trajectory. Vectors expressed in the $ECI$ frame can be transformed to the $LVLH$ frame through multiplication by the direction cosine matrix
\beq
    \mbf{C}_{LVLH-ECI} =
    \bbm
        \mbf{LVLH}^1_{ECI} & \mbf{LVLH}^2_{ECI} & \mbf{LVLH}^3_{ECI} 
    \ebm^\trans.
\eeq
For example, $\mbf{r}^{sp}_{LVLH} = \mbf{C}_{LVLH-ECI}\mbf{r}^{sp}_{ECI}$.

The position of the spacecraft with center of mass at point $s$ relative to its guidance position $g$, expressed in the $LVLH$ frame, is written as $\mbf{r}^{sg}_{LVLH}$ and is found through $\mbf{r}^{sg}_{LVLH} = \mbf{r}^{sp}_{LVLH}-\mbf{r}^{gp}_{LVLH}$. The velocity of the spacecraft relative to the guidance trajectory with respect to the $ECI$ frame and expressed in the $LVLH$ frame is written as $\mbf{v}^{sg/ECI}_{LVLH}$ and is obtained by differentiating the relative position with respect to the $LVLH$ frame using the transport theorem, giving
\beq
    \mbf{v}^{sg/ECI}_{LVLH} = \mbf{v}^{sp/LVLH}_{LVLH}-\mbf{v}^{gp/LVLH}_{LVLH} - \left(\mbs{\omega}_{LVLH}^{LVLH-ECI}\right)^\times\mbf{r}^{sg}_{LVLH},
\eeq
where $\mbf{v}^{sp/LVLH}_{LVLH} = \dot{\mbf{r}}^{sp}_{LVLH}$, $\mbf{v}^{gp/LVLH}_{LVLH} = \dot{\mbf{r}}^{gp}_{LVLH}$, 
\beq
    \mbs{\omega}_{LVLH-ECI} = \frac{\left({\mbf{r}^{gp}_{LVLH}}\right)^\times\mbf{v}^{gp/ECI}_{LVLH}}{{\mbf{r}^{gp}_{LVLH}}^\trans\mbf{r}^{gp}_{LVLH}}
\eeq
is the angular velocity of the $LVLH$ frame relative to the $ECI$ frame~\cite{Curtis3e}, and $\mbf{v}^{gp/ECI}_{LVLH} = \mbf{C}_{LVLH-ECI}\dot{\mbf{r}}^{gp}_{ECI}$.

\subsection{Drag Modulation}
The atmospheric drag acting on a spacecraft in LEO causes its orbit to decay until it ultimately reenters the Earth's atmosphere. The rate at which this decay occurs is related to the aerodynamic and mass properties of the spacecraft. A spacecraft that experiences a larger drag force relative to its mass will have its orbit decay more quickly. The amount of drag compared to the spacecraft's mass can be described by the ballistic coefficient, defined in this paper as 
	\beq
		C_b = \frac{C_dA_\text{ref}}{2m},
	\eeq
where $C_d$ is the drag coefficient of the spacecraft, $A_\text{ref}$ is the cross-sectional area and $m$ is the mass of the spacecraft. This definition of the ballistic coefficient is inverted compared to the more typical definition and contains an additional factor of one half. Using this definition, the drag acceleration becomes proportional to $C_b$ which results in dynamic equations that are more convenient for the purposes of estimation and control. The acceleration experienced by the spacecraft due to drag is written as
\beq
	a_d = \frac{\frac{1}{2}\rho v^2 C_d A_\text{ref}}{m},
\eeq
where $\rho$ is the density of the local atmosphere and $v$ is the velocity of the spacecraft relative to the atmosphere. This can be related to the ballistic coefficient by
\beq
	\label{eq:drag_acc}
	a_d = \rho v^2 C_b.
\eeq
If the ballistic coefficient is changed to have a value of 
\beq
\label{eq:Cb_tot}
C_b = C_{b,nom} +\Delta C_b, 
\eeq
then the acceleration becomes
\beq
\label{eq:acc_drag_deltaCb}
	a_d = \rho v^2 C_b = \rho v^2(C_{b,nom}+\Delta C_b) = \rho v^2 C_b + \rho v^2 \Delta C_b = a_{d,nom} + \Delta a_{d,\Delta C_b}.
\eeq
The relation in~\eqref{eq:acc_drag_deltaCb} demonstrates that if the ballistic coefficient of a spacecraft is changed from a nominal value by $\Delta C_b$, then the acceleration due to drag will change from the nominal value by $\Delta a_{d,\Delta C_b}$. A change in ballistic coefficient can therefore be used as a control input to alter the trajectory of the spacecraft by affecting the acceleration due to drag.

Multiple drag control devices are currently being developed. One example is the ExoBrake from NASA Ames~\cite{Murbach2010}. The ExoBrake consists of drag surface deployed behind the spacecraft and attached with rigid struts. The struts can be reeled in or out to deploy or collapse the drag surface, enabling the ballistic coefficient to be modulated. ExoBrakes are currently being tested on flights of the TechEdSat series of CubeSats. Another drag control device is the Drag Deorbit Device (D3) from the University of Florida, which augments the surface area through the use of retractable tapes~\cite{Guglielmo2019}. The proposed design of D3 aims to be able to adjust the area of a spacecraft by $0.5$~m$^2$.

\subsection{Atmospheric Density Prediction}
Predicting the path that a spacecraft will follow under the influence of atmospheric drag requires advance knowledge of the atmospheric density along that path in order to compute the drag force. High-fidelity models of the atmosphere, such as NRLMSISE-00~\cite{Picone2002} and Jacchia-type models~\cite{Bowman2008}, are used to predict the density that a spacecraft will encounter while accounting for variations in the atmospheric density due to latitude, longitude, altitude, time of day, day of year, as well as variations due to the space weather environment. While atmospheric density model inputs such as position and time are precisely known along a given trajectory, the space weather environment is difficult to predict accurately which results in a large source of uncertainty in the density.

The state of Earth's magnetic field as well as the activity of the Sun can cause the density of the upper atmosphere to vary by orders of magnitude. The effect of Earth's magnetic field is included in high-fidelity models through the $K_p$ and $a_p$ geomagnetic indices~\cite{Vallado2014}. The activity of the Sun is included in atmosphere models using a solar index such as $F10.7a$~\cite{Vallado2014}, which is measured by observing radio flux at a wavelength of $10.7$cm. While the effects of the geomagnetic index and solar flux are accounted for in high-fidelity atmosphere models, obtaining advance knowledge of atmospheric density from these models requires advance knowledge of the space weather environment, as described by these indices. 

Drag-modulated spacecraft trajectories may have a duration ranging from several weeks to many months, requiring predictions of density along the entire trajectory. However, predicting the space weather environment is challenging and forecasts lose their accuracy after projecting only a few days into the future~\cite{F10.7forecast, ApForecast}, leading to poor accuracy in density predictions and errors in propagating the spacecraft state~\cite{Vallado2014}. Additionally, even if perfect knowledge of the future space weather environment is available, the predictions made by these models still suffer from root mean square errors of up to $30\%$ over a trajectory~\cite{rhoden2000influence,picone2002nrlmsise,Doornbos2005}. This uncertainty in the atmospheric density is a primary factor limiting the accuracy of orbital predictions~\cite{Doornbos2006}. 

Due to inaccurate space weather forecasting and atmosphere modeling errors the atmospheric density, $\rho$, that the spacecraft experiences in-flight will differ from the predicted density, $\rho_{nom}$, leading to 
\beq
	\label{eq:rho_tot}
	\rho = \rho_{nom}+\Delta\rho.
\eeq
Similarly to~\eqref{eq:acc_drag_deltaCb}, this difference in density, $\Delta\rho$, will lead to a change in acceleration from the nominal of
\beq
a_d = \rho v^2C_b = (\rho_{nom} + \Delta\rho)v^2C_b = \rho_{nom}v^2C_b + \Delta\rho v^2C_b = a_{d,nom}+\Delta a_{d,\Delta\rho}.
\eeq	
Likewise, this change in acceleration compared to the nominal will cause the spacecraft to deviate from the intended trajectory. However, if the value of $\Delta\rho$ can be estimated, as proposed in this paper, it can be used to generate drag-modulation control inputs that compensate for the difference in density to prevent the spacecraft from departing from the desired trajectory. 

\section{Methodology}
\label{sec:Methodology}

This section presents the methodology of the proposed estimation and control framework for drag-modulated reentry targeting. The guidance trajectory generation algorithm is first presented, which is primarily based on the work of~\cite{Omar2017conf,Omar2019}, and features a minor contribution on how to choose the nominal ballistic coefficients to increase the authority of the tracking controller. The relevant equations of motion used by the proposed estimation and control approaches are then presented, followed by the EKF-based density error estimation method and the MPC-based tracking controller.

\subsection{Guidance Trajectory Generation}
\label{sec:GuidanceGeneration}
Guidance trajectories are generated with a simplified method based on work by Omar~\cite{Omar2017conf,Omar2019} from the initial condition of the spacecraft to the desired latitude and longitude at the entry interface altitude. The guidance scheme considers a single modulation of the ballistic coefficient from an initial value of $C_{b1}$ to a final value of $C_{b2}$ that occurs at a time $t_\text{swap}.$ The guidance trajectory generation problem then consists of finding the values of these guidance parameters that yield a trajectory with the desired latitude and longitude at the entry interface altitude.

Omar's method involves deriving simplified analytical relationships between the guidance parameters and the entry interface location. Given an orbital decay trajectory, these equations predict how the time of flight and number of orbits made during the trajectory will change based on a change in the guidance parameters. These equations are used to compute new values of the guidance parameters that correct for the targeting errors of a given trajectory, enabling an initial guess of the guidance trajectory to be iteratively refined until proper targeting is achieved. 

The iteration involves distinct latitude and longitude targeting steps. In the latitude targeting step, $t_\text{swap}$ is altered to change the total number of orbits made during the trajectory, affecting the final latitude. After $t_\text{swap}$ is obtained, $C_{b1}$ and $C_{b2}$ are computed to change the time of flight, which alters the final longitude as the orbit precesses and the Earth rotates. The orbital decay of a spacecraft following this scheme is shown in Fig.~\ref{fig:Cb_swap}. A detailed summary of Omar's algorithm adapted for use in this work is presented in the Appendix for completeness.

\begin{figure}[t!]
	\centering
		\includegraphics[width=0.6\linewidth]{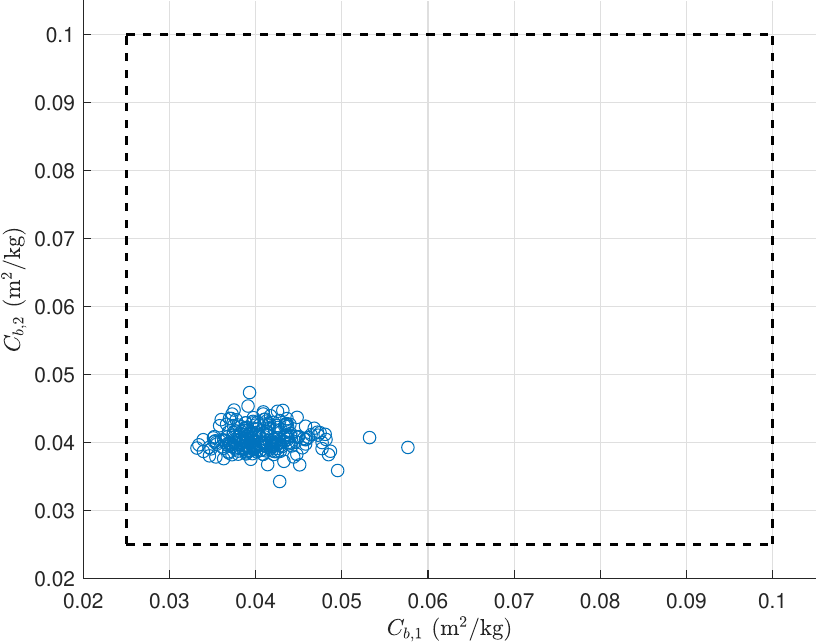} 
	\caption{Nominal ballistic coefficients for 250 trajectories with randomly sampled initial conditions and desired entry interface locations compared to the feasible range.}
	\label{fig:Cb_max_ctrbl}
\end{figure}

If the spacecraft has sufficient control authority and the trajectory begins at a sufficiently high altitude, there may be many valid combinations of the guidance parameters that yield proper reentry targeting. One approach involves selecting the $t_\text{swap}$ that targets the desired latitude at entry interface with the smallest residual longitude error \cite{Omar2017conf} to be corrected in the longitude targeting step. Another approach is to choose the $t_\text{swap}$ that provides the largest amount of controllability margin about the desired longitude~\cite{Omar2017}. However, this approach tends to produce guidance ballistic coefficients $C_{b1}$ and $C_{b2}$ toward the bottom of the feasible range of ballistic coefficients, limiting the amount of control authority in one direction. To illustrate this, the approach from~\cite{Omar2017} is used to generate 250 guidance trajectories with a feasible ballistic coefficient range of $C_{b,\text{min}} = 0.025$~m$^2$/kg to $C_{b,\text{max}} = 0.1$~m$^2$/kg. As shown in Fig.~\ref{fig:Cb_max_ctrbl}, the resulting guidance ballistic coefficients are clustered around $C_b = 2\frac{C_{b,\text{min}}C_{b,\text{max}}}{C_{b,\text{min}}+C_{b,\text{max}}} = 0.04 $~m$^2$/kg. With this value of the guidance ballistic coefficients, the controller is able to increase the ballistic coefficient, and therefore the drag, by 150\%. However, the drag can only be decreased by 37.5\%, leading to far more control authority to increase drag compared to reducing drag. This degrades the ability of the controller to track trajectories when the density is larger than expected. Omar~\cite{Omar2019} addresses this by only using a portion of the feasible range of $C_b$ when generating the guidance in order to reserve some of the range of $C_b$ for tracking. However, more control authority can be preserved if the guidance trajectory is generated in such a way to produce guidance ballistic coefficients in the middle of the feasible range. 

A minor contribution of this paper is to choose a ballistic coefficient in the middle of the feasible range, $C_{b,\text{mid}}$ such that a spacecraft with this ballistic coefficient can reduce or increase the drag acting on it by the same factor, that is, 
\beq
    \frac{C_{b,\text{mid}}}{C_{b,\text{min}}} = \frac{C_{b,\text{max}}}{C_{b,\text{mid}}},
\eeq
The desired ballistic coefficient for the guidance is then
\beq
    C_{b,\text{mid}} = \sqrt{C_{b,\text{min}}C_{b,\text{max}}}.
\eeq
For each feasible $t_\text{swap}$, the $C_{b1}$ and $C_{b2}$ that produce the desired entry interface location are computed. The $t_\text{swap}$ is then selected according to
\beq
    \underset{t_\text{swap}}{\min} \norm{\bbm C_{b1} - C_{b,\text{mid}}\\C_{b2} - C_{b,\text{mid}}  \ebm}.
\eeq
The results presented in Section~\ref{sec:Results} demonstrate that this choice of $t_\text{swap}$ produces nominal ballistic coefficients that are closer to the center of the feasible range compared to Fig.~\ref{fig:Cb_max_ctrbl}.

\begin{figure}[t!]
	\centering
\subfloat[]{		                  \includegraphics[width=0.48\linewidth]{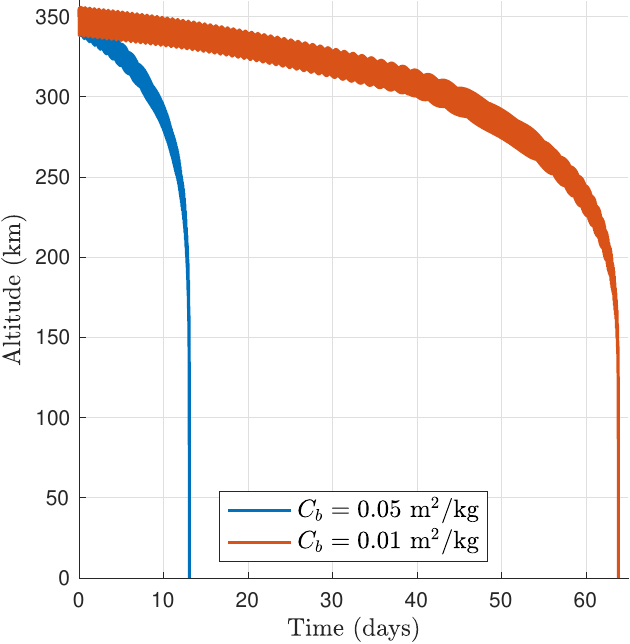}}
\subfloat[]{		                  \includegraphics[width=0.48\linewidth]{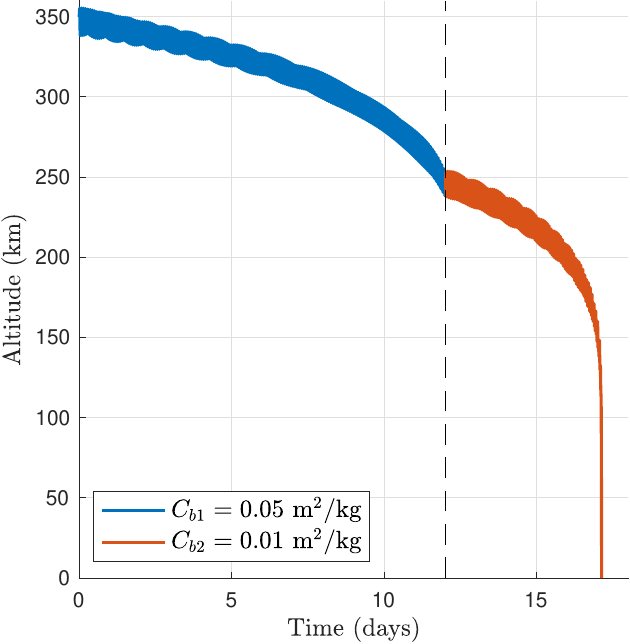}}
	\caption{Orbital decays of spacecraft with (a) constant ballistic coefficients and (b) a change in ballistic coefficient at a $t_{swap}$ of approximately 12 days.}
	\label{fig:Cb_swap}
\end{figure}

\subsection{Equations of Motion}
A linear model is required for both the EKF as well as the proposed MPC approach. It is assumed, at all times, the spacecraft is in a nearly circular orbit and that the distance between the spacecraft and the guidance position is small compared to the radius of the Earth. These assumptions enable the use of the linear model of relative motion developed by Schweighart and Sedwick~\cite{Schweighart2002}. This model is similar to the Clohessy-Wiltshire equations~\cite{Curtis3e} but differs in that it captures some effects of the $J2$ perturbation.

In the linear model, the motion in the cross-track direction is decoupled from the motion in the radial and in-track directions. Cross-track errors are uncontrollable because the control input is based on changes in drag, which acts primarily opposite to the in-track direction, which is not coupled to the cross-track direction in the model. However, if the initial cross-track error is small and the error in the radial and in-track directions are kept sufficiently small, the orbit of the spacecraft will precess at a similar rate to the guidance orbit which prevents excessive growth in the cross-track errors. 

Because the cross-track motion is uncontrolled, it is omitted from the state vector of the spacecraft for the purpose of estimation and control. The state vector, $\mbf{x}_{dyn}$, is composed of the relative position and velocity components in the radial and in-track directions (i.e., the first and second components of $\mbf{r}^{sg}_{LVLH}$ and $\mbf{v}^{sg/ECI}_{LVLH}$) according to
\beq
\mbf{x}_{dyn}  = \bbm \mbf{r}^{sg}_{LVLH,1} \\ \mbf{r}^{sg}_{LVLH,2} \\ \mbf{v}^{sg/ECI}_{LVLH,1} \\ \mbf{v}^{sg/ECI}_{LVLH,2} \ebm.
\eeq
Because this position and velocity are relative to the guidance trajectory, this state vector also represents the tracking errors. In the absence of density prediction error and control input, the rate of change of the state vector is found through
\beq
    \label{eq:SS_auto}
	\dot{\mbf{x}}_{dyn}(t) = \mbf{A}(t)\mbf{x}_{dyn}(t), 
\eeq 
where the matrix $\mbf{A}$ is computed using the equations derived by Schweighart and Sedwick \cite{Schweighart2002}, giving 
\beq
	\mbf{A}(t) = 
	\bbm
		0&0&1&0 \\
		0&0&0&1 \\ 
		b(t)&0&0&d(t) \\ 
		0&0&-d(t)&0
	\ebm.
\eeq
The entries of this matrix are given by
\beqarray
n(t) &=& \sqrt{\frac{\mu}{a(t)^3}}, \\ c(t) &=& \sqrt{1 + \frac{3J_2R_e^2}{8a(t)^2}[1+3\cos{2i(t)}]}, \\ d(t) &=& 2n(t)c(t), \\ b(t) &=& (5c(t)^2 - 2)n(t)^2,
\eeqarray
where $\mu$ is the standard gravitational parameter, $J_2$ is the constant for the gravitational J2 perturbation~\cite{Vallado2013}, and $R_e$ is the average radius of the Earth. The semi-major axis of the guidance orbit, $a(t)$, as well as the inclination, $i(t)$, both vary in time due to drag, leading to the time-variant nature of $\mbf{A}(t)$.

If the spacecraft is operating at the nominal ballistic coefficient specified by the guidance and there is no density error, then there will be no differential drag between the guidance and the spacecraft. In this situation, there is no relative acceleration due to drag and the model in~\eqref{eq:SS_auto} can be used to represent the motion between the spacecraft and the guidance. However, in the presence of a control input $\Delta C_b$ and a density prediction error $\Delta \rho$, the acceleration due to drag becomes
\begin{align}
    a_d &= \rho v^2 C_b = \left(\rho_{\text{nom}}+\Delta \rho_\text{nom}\right) v^2 \left(C_{b,\text{nom}} + \Delta C_b\right) \nonumber \\
     &=\rho_\text{nom} v^2 C_{b,\text{nom}} + \Delta\rho v^2 C_{b,\text{nom}} + \rho_\text{nom} v^2 \Delta C_{b} + \Delta\rho v^2 \Delta C_{b} \nonumber \\
     &= a_{d,\text{nom}} + \Delta a_{d},
\end{align}
where the drag along the guidance trajectory is $a_{d,\text{nom}} = \rho_\text{nom} v^2 C_{b,\text{nom}}$ and 
\beq
    \Delta a_d =  \Delta\rho v^2 C_{b,\text{nom}} + \rho_\text{nom} v^2 \Delta C_{b} + \Delta\rho v^2 \Delta C_{b}
\eeq
is the differential drag acceleration due to control input as well as density error. Because drag acts primarily in the direction opposite the velocity, the differential drag acceleration is added to the linear model as an acceleration in the negative in-track direction yielding
\beq
    \label{eq:SS_drag}
    \dot{\mbf{x}}_{\text{dyn}}(t) = \mbf{A}(t)\mbf{x}_\text{dyn}(t) + \bbm \mbf{0}\\-\Delta a_d(t) \ebm , 
\eeq
which is a model that includes the effect of density errors and control inputs. 

\subsection{Density Prediction Error Estimation}
A discrete-time EKF is used to estimate the density prediction error, $\Delta\rho$, as well as the radial and along-track position and velocity of the spacecraft relative to its guidance trajectory as a byproduct.  It is assumed that a GPS receiver is used to provide measurements of the position and velocity of the spacecraft relative to the guidance position and velocity. An estimate of the density error and the spacecraft state is made at the beginning of each time step, which is used within the MPC strategy.

\subsubsection{Process Model}
To construct the process model for the filter, first it is assumed that the density difference at each point in time is a fraction, $c_{\Delta\rho}$, of the nominal density and is given by 
\beq
	\label{eq:c_def}
	\Delta\rho(t) = c_{\Delta\rho}(t) \rho_{nom}(t).
\eeq
The state of the spacecraft is then augmented with the fraction $c_{\Delta\rho}$, which represents a normalized density prediction error, to form the state of the EKF, given by
\beq
    \label{eq:KF_state}
	\mbf{x}_{KF} = \bbm \mbf{x}_{dyn} \\ c_{\Delta\rho} \ebm.
\eeq 
The subscript $KF$ is included in the definition of this state to indicate that it is the state associated with the EKF. The process that governs $\mbf{x}_\text{dyn}$ is given by~\eqref{eq:SS_drag}. However, the process that governs $c_{\Delta\rho}$ can not be easily represented as a linear function of the spacecraft state, $\mbf{x}_\text{dyn}$. The evolution of $c_{\Delta\rho}$ is the result of complex interactions between the activity of the sun, the state of Earth's magnetic field, the altitude of the spacecraft and many more factors that affect the density of Earth's atmosphere. Due to the intractability of developing a linear process model, the evolution of $c_{\Delta\rho}$ is modeled as a random walk in discrete time. Random walk processes have been used in the literature to estimate atmospheric density uncertainty~\cite{schiemenz2019least,giraldo2024precision,hinks2010simultaneous}, which provides additional confidence in this modeling choice.

The continuous-time dynamics in~\eqref{eq:SS_drag} are factored to group the $\Delta \rho$ terms which is then substituted by~\eqref{eq:c_def}. This allows the time derivative of~\eqref{eq:KF_state} to be written as  
\beq    
\label{eq:proc_model}
\dot{\mbf{x}}_{KF}(t) = \mbf{A}_{KF}(t)\mbf{x}_{KF}(t) + \mbf{B}_{KF}(t) \Delta C_{b}(t),
\eeq
where
\beq
    \mbf{A}_{KF}(t) = \bbm  0&0&1&0 & 0\\
		0&0&0&1 & 0 \\ 
		b(t)&0&0&d(t) & 0 \\ 
		0&0&-d(t)&0 & -\rho_\text{nom}(t) v(t)^2\left(C_{b,\text{nom}}(t) + \Delta C_b (t)\right) \\ 0&0&0&0 & 0 \ebm, \hspace{20pt} 
\mbf{B}_{KF}(t) = \bbm 0\\0\\0 \\ -\rho_\text{nom}(t)v(t)^2 \\ 0 \ebm.
\eeq

The guidance trajectory and the trajectory of the spacecraft are discretized into time steps of $\Delta T$ in length. In order to propagate the spacecraft state between time steps the model in~\eqref{eq:proc_model} is discretized in time assuming a zero-order hold (ZOH) and a sample time of $\Delta T$. Process noise is added to obtain
\beq
\label{eq:disc_proc}
\mbf{x}_{KF,k+1} = \mbf{A}_{KF,k}\mbf{x}_{KF,k} + \mbf{B}_{KF,k} \Delta C_{b,k} + \mbf{L}\mbf{w}_k,
\eeq
where $\mbf{L} = \bbm \mbf{0}_{3 \times 3} & \mbf{1}_{3 \times 3} \ebm^\trans$. The ZOH approximates all time-varying parameters in the model as constant thoughout each time step. The discrete-time model is recomputed at each time step using the values of the parameters at the beginning for the time step (e.g., $v(t) = v(t_k)$ for $t\in\left(t_k, t_{k+1} \right]$). The ZOH also accurately reflects that MPC holds the control input $\Delta C_b$ constant over each time step. 
The process noise, $\mbf{w}_k \in \mathbb{R}^3$, with a covariance of $\mbf{W} = \mathbb{E}\{\mbf{w}_k \mbf{w}_k^\trans\}$, is added to the velocity states as well as $c_{\Delta\rho}$ to account for nonlinearities in the dynamics that are neglected by the model and to enable the estimate of $c_{\Delta\rho}$ to change over time.  

\subsubsection{Time Update}
To obtain an estimate of the state at time-step $k$, first the previous estimate is propagated forward in time using the process model in~\eqref{eq:disc_proc}, resulting in 
\beq
		\hat{\mbf{x}}^-_{KF,k} = \mbf{A}_{KF,k-1}\hat{\mbf{x}}^+_{KF,k-1}+\mbf{B}_{KF,k-1}\Delta\ C_{b,k-1}.
\eeq	
The covariance of the state estimation error is propagated in time using
\beq
 	\mbf{P}^-_{k} = \mbf{A}_{KF,k-1}\mbf{P}^+_{k-1}\mbf{A}_{KF,k-1}^\trans + \mbf{L}\mbf{WL}^\trans.
\eeq

\subsubsection{Measurement Update}
After the time update, the most recent measurement, $\mbf{y}_k$, is input to the filter to provide an updated state estimate. This measurement is one of position and velocity of the spacecraft relative to the guidance trajectory at time $t = t_k$  and is obtained from a GPS receiver and assumed to be unbiased. The filter uses the measurement model
\beq
    \label{eq:meas}
	\mbf{y}_k = \mbf{H}\mbf{x}_{KF,k} + \mbf{v}_k,
\eeq
where $\mbf{H} = \bbm \mbf{1}_{4 \times 4} & \mbf{0}_{4 \times 1} \ebm$ maps the state vector to relative position and velocity, while $\mbf{V} = \mathbb{E}\{\mbf{v}_k\mbf{v}_k^\trans\}$ is the GPS velocity and position measurement error covariance. The updated state estimate, including the measurement, is then found by
\beq
	\hat{\mbf{x}}^+_{KF,k} = \hat{\mbf{x}}^-_{KF,k} + \mbf{P}^-_k\mbf{H}^\trans(\mbf{H}\mbf{P}^-_k\mbf{H}^\trans +\mbf{V})(\mbf{y}_k - \mbf{H}\hat{\mbf{x}}^-_{KF,k}).
\eeq
The updated covariance of the state estimation error is then computed according to
\beq
	\mbf{P}^+_k = \mbf{P}^-_k - \mbf{P}^-_k\mbf{H}^\trans(\mbf{H}\mbf{P}^-_k\mbf{H}^\trans +\mbf{V})^{-1}\mbf{H}\mbf{P}^-_k.
\eeq

\subsubsection[Extracting Estimates of Delta-rho]{Extracting Estimates of $\Delta\rho$}
An estimate of the value of $\Delta\rho$ at time-step $k$ is extracted from the filter by multiplying the nominal value of the density by the estimate of $c_{\Delta\rho}$ giving
\beq   
    \label{eq:Delta_rho_est}
	\Delta\hat{\rho}^+_k = \hat{c}_{\Delta\rho,k}^+\rho_{nom}(t_k).
\eeq
Because of the ZOH, this estimate of $\Delta\rho$ can be thought of as an effective density error over an entire time step rather than an instantaneous density error. This estimate of $\Delta\rho$ is then used as a predicted disturbance in the MPC formulation to compensate for the effect of density prediction errors. 

\subsection{Model Predictive Control}
In order to guide the spacecraft along the guidance trajectory and to the desired location at the entry interface altitude, the spacecraft state, as estimated by the EKF, is used as feedback within a MPC framework~\cite{Hayes2022}. In this framework, the estimated density prediction error is included in the MPC calculation as a predicted disturbance in order to compensate for differences between the density used in generating the guidance trajectory and the density encountered by the spacecraft in flight. 

The continuous-time dynamic model in~\eqref{eq:SS_drag} is factored to group the $\Delta C_b$ terms. Grouping these terms results in
\beq
    \label{eq:CT_MPC_dyn}
    \dot{\mbf{x}}_\text{dyn}(t) = \mbf{A}(t)\mbf{x}_\text{dyn}(t) + \mbf{B}_{\text{MPC},1}(t)\Delta C_b(t)+\mbf{B}_{\text{MPC},2}(t)\Delta\rho(t),
\eeq
where
\beq
    \mbf{B}_{\text{MPC},1}(t) = \bbm 0 \\ 0 \\ 0  \\ -\left(\rho_\text{nom}(t)+\Delta\rho(t) \right)v^2(t) \ebm,  \hspace{20pt} \mbf{B}_{\text{MPC},2}(t) = \bbm  0 \\ 0 \\ 0 \\ -v^2(t)C_{b\text{nom}}(t) \ebm.
\eeq
This model describes the motion of the spacecraft relative to the guidance and accounts for the effects of density prediction errors $\Delta\rho$ and control inputs $\Delta C_b$.

The guidance trajectory and the trajectory of the spacecraft are divided into discrete time steps of $\Delta T$ in length that align with the time steps of the EKF. The model given by~\eqref{eq:CT_MPC_dyn} is discretized in time, again using a ZOH on all of the time-varying parameters that appear in the model and a sample time of $\Delta T$, resulting in 
\beq   
    \label{eq:DT_MPC_dyn}
    \mbf{x}_{\text{dyn},k+1} = \mbf{A}_k\mbf{x}_{\text{dyn},k} + \mbf{B}_{\text{MPC},1,k}\mbf{x}_{\text{dyn},k} + \mbf{B}_{\text{MPC},2,k}\Delta\rho_k.
\eeq
As a result of the ZOH, changes in the control input only occur at the beginning of each time step resulting in predictable actuation times and limited actuator usage. Holding the control input constant over each time step is also consistent with how the control input is modeled in the EKF. 

Beginning at a particular time, the model in~\eqref{eq:DT_MPC_dyn} is used to propagate the spacecraft state $N$ time steps into the future beginning at the current time $t$ and initialized using the current estimate of the spacecraft state, extracted from $\hat{\mbf{x}}^+_{KF,k}$. The period of time spanned by the $N$ time steps is the prediction horizon.  The predicted state $k$ steps ahead of the current time is $\mbf{x}_{\text{dyn},k|t} = \mbf{x}_{\text{dyn}}\left(t + k\Delta T\right)$ for $k\in [0,N]$. The state trajectory throughout the prediction horizon is a function of the control input at each time step $\Delta C_{b,k|t} = \Delta C_b(t + k\Delta T)$ and the disturbance $\Delta \rho_{k|t} = \Delta\rho(t + k\Delta T)$ for $k\in [0,N]$. The estimate obtained from~\eqref{eq:Delta_rho_est} is used as a prediction of the disturbance. 

Previous work~\cite{Hayes2022} used a linear time-invariant (LTI) model where the discrete-time dynamics in~\eqref{eq:DT_MPC_dyn} are computed using $\rho_{\text{nom},0|t}$ and $\Delta\rho_{0|k}$ and assumed constant over the prediction horizon. Because the nominal density and, to a lesser extent, the spacecraft velocity are time-varying, an LTV model of the spacecraft dynamics is used, where $\mbf{A}_k$, $\mbf{B}_{\text{MPC},1,k}$, $\mbf{B}_{\text{MPC},2,k}$ are computed for each time step in the prediction horizon. Additionally, the density prediction error appears both in the state space matrices and as a disturbance. If the system were assumed to be LTI, then $\rho_{\text{nom},0|t}$ and $\Delta \rho_{0|t}$ would be used to compute the state-space matrices. This would be inconsistent with the use of $\rho_{\text{nom},k|t}$ and $\Delta\rho_{k|t}$, which are time-varying nominal density and density prediction error terms. The LTV formulation ensures that the dynamics matrices are computed using values that are consistent in timing with the density prediction error. 

When choosing a cost function for MPC to minimize, it is important to consider the practical needs of the mission. Because the spacecraft state is expressed as a position and velocity relative to the guidance trajectory, a feedback controller should drive these states to zero in order to drive the spacecraft to the guidance trajectory. This motivates the inclusion of a state cost that penalizes how far the spacecraft is from the guidance during the prediction horizon. The cost function should also penalize the use of control input as was done in previous work~\cite{Hayes2022}, where changes in ballistic coefficient compared to the nominal were penalized. Similarly, Omar~\cite{Omar2019} penalized changes in the ballistic coefficient in order to control when saturation of the drag device occurred. However, the control input is a change in ballistic coefficient from the nominal and no energy is required for a drag device to hold a particular ballistic coefficient. Energy is only required to run the actuator to change the ballistic coefficient. Therefore, because MPC already explicitly accounts for the effect of saturation, it is more useful to penalize changes in the control input between time steps in order to reduce excessive actuation.

Motivated by the prior work of~\cite{Omar2019,Hayes2022} and the desire to penalize only a change in control input between time steps, the proposed MPC cost function is
\beq    
    \label{eq:MPC_cost}
    J\left(\mbf{U}\right) = \sum_{k = 0}^{N-1} \mbf{x}_{k|t}^\trans \mbf{Q}_c\mbf{x}_{k|t}  + \mbf{x}_{N|t}^\trans \mbf{Q}_{c,N}\mbf{x}_{N|t}+\sum_{k=0}^{N-2} R_c(\Delta C_{b,k+1}-\Delta C_{b,k})^2,
\eeq
where $\mbf{U} = \bbm \Delta C_{b,0|t} & \cdots & \Delta C_{b,N-1|t}\ebm^\trans$ is a vector of all the control inputs throughout the prediction horizon. The relative cost of position and velocity errors in each direction throughout the prediction horizon, not including the state at the end of the prediction horizon, is weighted by $\mbf{Q}_c$ where as the terminal state error cost is weighted by $\mbf{Q}_{c,N}$. The relative cost associated with changes in control input is weighted by $R_c$. If the tracking errors are too large, they can be reduced at the expense of more changes in control input by increasing the diagonal entries of $\mbf{Q}_c$ and $\mbf{Q}_{c,N}$ compared to $R_c$. Conversely, if the controller is commanding excessive changes in ballistic coefficient resulting in too much use of the control actuators, $R_c$ can be increased to increase the penalty associated with changes in the control input resulting in less actuation at the expense of larger tracking errors.

The cost in~\eqref{eq:MPC_cost} is minimized over $\mbf{U}$ to obtain the optimal sequence of ballistic coefficient changes throughout the prediction horizon. This optimization is constrained such that the commanded ballistic coefficients lie in a feasible range attainable by the spacecraft, $C_b \in [C_{b,\text{min}},C_{b,\text{max}}]$. Because the control inputs are changes in the ballistic coefficient compared to the guidance ballistic coefficient, bounds on $\Delta C_{b,k|t}$ vary in time and are computed based on the guidance ballistic coefficient. For a guidance trajectory with a ballistic coefficient of $C_{b,g}(t)$, the maximum allowable change in ballistic coefficient at any given time is
\beq
    \Delta C_{b,\text{max}}(t) = C_{b,\text{max}}-C_{b,g}(t).
\eeq
Similarly, the most negative allowable change in ballistic coefficient is 
\beq
    \Delta C_{b,\text{min}}(t) = C_{b,\text{min}}-C_{b,g}(t). 
\eeq
These bounds are also applicable to the discrete time optimization of~\eqref{eq:MPC_cost} if the guidance ballistic coefficient is constant throughout a time step. However, if the guidance ballistic coefficient changes during a time step, a conservative bound must be used to ensure the control limits are not violated. The control inputs can then be constrained by
\beq
    \label{eq:DeltaCb_constraint}
 \Delta C_{b,\text{min},k|t} \leq \Delta C_{b,k|t} \leq \Delta C_{b,\text{max},k|t}
\eeq
to ensure that the commanded ballistic coefficient always lies in the feasible range of $C_b \in [C_{b,\text{min}},C_{b,\text{max}}]$.

The optimization consisting of the minimization of~\eqref{eq:MPC_cost} subject to the constraints of the control input given by~\eqref{eq:DeltaCb_constraint} is formulated as a quadratic program, which is convex and can be solved quickly and efficiently. For convenience, the built-in MATLAB quadratic program solver \textit{quadprog} is used to solve the optimization problem in this work, however other solvers such as OSQP~\cite{stellato2020osqp} or custom code generated by CVXgen~\cite{mattingley2012cvxgen} are suitable for real-time use in an embedded application such as onboard a spacecraft. 

The solution of the optimization provides a sequence of control inputs to apply over each time step in the prediction horizon. However, only the the first in the sequence is applied by the spacecraft. At the end of the current time step, the EKF produces a new estimate of the spacecraft state and the density error, and the optimization is repeated with these new estimates to produce a new sequence of control inputs.

In order to improve tracking performance without increasing the number of control actuations or increasing the size and complexity of the quadratic program, each time step is divided into a number of substeps, $N_{ss}$. The dynamics are then discretized over each substep. Because the sample time of the substeps is smaller than that of the time step, the errors due to the zero-order-hold approximation of the time-varying parameters in the dynamic model are reduced. The number of actuations does not change because the control is held constant across all substeps in the time step. 

\section{Simulations}
\label{sec:Simulations}
\begin{figure}[t!]
	\centering
		\includegraphics[width=0.85\linewidth]{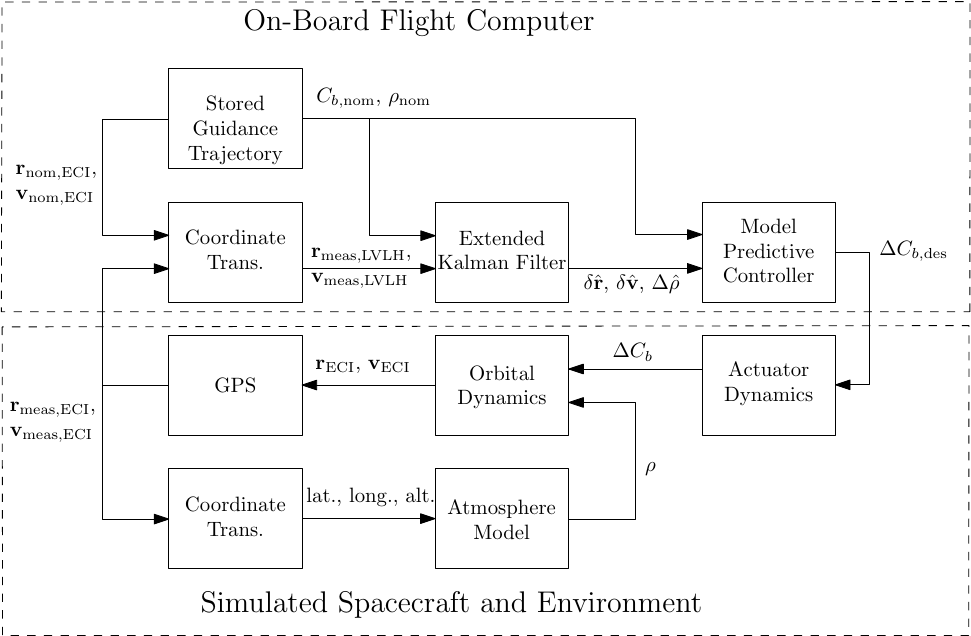} 
	\caption{Block diagram depicting the simulation of estimation and control algorithms and the spacecraft dynamics.}
	\label{fig:EKF_MPC_blk_diag}
\end{figure}

To evaluate the performance of the proposed estimation and control framework under realistic flight conditions, a Monte-Carlo campaign consisting of 1000 runs is conducted. Each run generates a guidance trajectory beginning in a random near-circular orbit at approximately $375$ km in altitude and ending at a random latitude and longitude at the von Karman altitude of $100$~km. Although the altitude of entry interface is not precisely defined in the literature~\cite{gallais2007atmospheric}, $100$~km is chosen in this work to align with prior studies on drag-based deorbit trajectories~\cite{Virgili2015,fedele2021precise,fedele2021aerodynamic}. The estimation and control framework is then simulated in a custom simulation environment beginning at a dispersed initial position and velocity and in the presence of atmospheric density errors, gravitational modeling errors and measurement noise. The simulation continues until the spacecraft reaches $100$ km in altitude and the distance by which the spacecraft missed the desired reentry location is recorded. Should the tracking spacecraft drift away from the guidance trajectory by more than $100$~km at any point in time, the simulation is terminated and the tracking is considered to have failed. The simulation environment, along with the EKF and MPC, are depicted in Fig.~\ref{fig:EKF_MPC_blk_diag}. For comparative purposes, the estimation and control framework is simulated twice for each trajectory. One simulation makes use of the estimated density error within MPC, while the other does not. 

\subsection{Environmental Modeling}
A custom simulation environment is developed in MATLAB leveraging the high-fidelity models included in the Aerospace Toolbox~\cite{aero_tbx}. These include models of Earth's gravity and atmosphere, which are used to calculate the different forces that influence the motion of the spacecraft. The dynamics of the spacecraft are propagated using the variable-step \textit{ode113} solver. This custom simulation environment has been validated against multiple simulation test cases provided by the NASA Engineering and Safety Center~\cite{Jackson2013}.

\subsubsection{Nominal Models}
The Earth is modeled as an oblate spheroid using the WGS84~\cite{Slater1997} ellipsoid. The altitude of the spacecraft is then calculated as the height of the spacecraft above this ellipsoid. The Earth is assumed to rotate about the $\underrightarrow{ECI}^3$ direction resulting in an angular velocity of the $ECEF$ frame relative to the $ECI$ frame of $\mbs{\omega}^{ECEF-ECI}_{ECI} = \bbm 0 & 0 & 7.292115\times 10 ^{-5} \ebm^\trans$~rad/s. The orientation of the Earth at the initial time is obtained from the International Astronomical Union - 2000/2005 reference system as implemented in the MATLAB function \textit{dcmeci2ecef}~\cite{aero_tbx} and the nutation and precession is neglected.

The gravitational acceleration acting on the spacecraft is modeled using the EGM2008~\cite{Pavlis2012} model as implemented in the MATLAB function \textit{gravitysphericalharmonic}~\cite{aero_tbx}. This model is used with degree and order 15 to generate nominal values of the gravitational acceleration. 

The drag acceleration acting on the spacecraft is computed according to~\eqref{eq:drag_acc}. The velocity of the spacecraft $v$ is the magnitude of the spacecraft relative to the atmosphere. The atmosphere is assumed to be motionless relative to the Earth, effectively neglecting all winds. At the location of the spacecraft, the atmosphere therefore has a velocity relative to the center of the Earth with respect to the $ECI$ frame of 
\beq
    \mbf{v}^{\text{atmo},p/ECI}_{ECI} = \left(\mbs{\omega}^{ECEF-ECI}_{ECI}\right)^\times\mbf{r}^{sp}_{ECI}.
\eeq
The velocity of the spacecraft relative to the atmosphere is then
\beq
    \mbf{v}^{s,\text{atmo}/ECI}_{ECI} = \mbf{v}^{sp/ECI}_{ECI} - \mbf{v}^{\text{atmo},p/ECI}_{ECI}.
\eeq
The magnitude of this relative velocity is used in computing the magnitude of the drag acceleration, whereas the direction of the relative velocity is the direction in which the drag acceleration acts.

The density of the atmosphere is obtained from the NRLMSISE00~\cite{Picone2002} atmosphere model as implemented in \textit{atmosnrlmsise00}~\cite{aero_tbx}.To generate nominal values of the density, the 81-day centered average of the $F10.7$ solar flux is used in place of the daily observed values. In place of the $A_p$ geomagnetic index for the current time, the average of eight 3-hour $A_p$ observations from 36 to 57 hours before the current time is used. These averages are used to represent the imperfect knowledge of future space weather indices that would be available in the form of forecasts when predicting a nominal density profile. 

\subsubsection{Dispersed Models}
When generating guidance trajecories, the nominal environmental models are used. However, in order to evaluate the performance of the estimation and control methods in the presence of modeling errors and uncertainties, dispersed environmental models are used in the tracking simulations.

In order to represent modeling error and uncertainty in the gravitational acceleration, the EGM2008~\cite{Pavlis2012} model is dispersed by using a higher degree and order of 40 compared to the nominal gravity model which produces differences in the effect of gravity between the guidance trajectory and the tracking simulation. 

In the tracking simulations, real historical observations~\cite{celestrak} of the $F10.7$ solar flux and $A_p$ geomagnetic index are used when evaluating the atmosphere model in order to provide realistic variations in the density. The discrepancy in the space weather indices between the guidance trajectory and the tracking simulation is used to produce density prediction errors representative of those that result from the errors in space weather forecasting.

There are other sources of uncertainty in the drag force, such as error in the atmosphere model itself, as well as uncertainty in the area and drag coefficient of the drag device. For the tracking simulations, a constant multiplier $c_\text{drag}$ is added to the drag acceleration to represent these additional sources of drag error. This multiplier is randomly sampled between individual simulations and held constant for the duration of each trajectory. 

If the density prediction error is too large, the spacecraft will not be able to change the ballistic coefficient enough to compensate and produce the nominal drag acceleration required to stay on the guidance trajectory. In this case, the drag device will saturate and the spacecraft will begin to drift away from the guidance. If this condition persists for long enough, the spacecraft may move so far from the guidance that it will be unable to recover. In this situation a new guidance must be generated and, depending on how late in the trajectory this occurs, there is no guarantee that the spacecraft will be able to reach the desired reentry location. Because guidance generation is not a contribution of this paper, these situations are considered as failures to track the trajectory and no attempt is made to generate and track a new trajectory. It is assumed that in a real-life scenario, an attempt would be made to produce and track a new trajectory. 

Previous work~\cite{Hayes2023} applied modern direct methods for trajectory optimization to determine whether it is possible for the spacecraft to return in finite time, or if a new trajectory must be generated. Other work~\cite{Omar2019} avoided this issue altogether by reserving control authority during guidance generation and never applying density dispersions that are large enough such that the spacecraft is not physically capable of compensating. However, in reality, there is no guarantee that the density dispersions will be small enough to be within the capability of the spacecraft to compensate. 

\subsection{Monte Carlo Parameters}
\label{sec:MC_param}

The sampled parameters used in generating the guidance trajectory are listed in Table~\ref{table:traj_param}. For each individual simulation, the initial condition of the guidance trajectory is generated by sampling the orbital elements of the initial orbit at a random epoch obtained from uniform distributions. All of these orbits begin at an altitude of approximately $375$ km by choosing an initial semi-major axis of $6853$~km, which was found to be sufficiently high to provide enough longitude controllability to target any point below the inclination of the spacecraft orbit, regardless of the state of the atmosphere. The desired latitude and longitude at entry interface are also sampled and the entry interface altitude is chosen to be $100$ km. 

\begin{table}[!t]
\centering
\begin{tabular}{cccc}
    \toprule
    Parameter & Minimum  & Maximum & Units \\
     \midrule
    Epoch & 0:00:00 1/1/1958& 0:00:00 1/1/2020& Time (UTC)  \\
    Eccentricity & $10^{-6}$&$10^{-3}$  &\\
    Inclination & $0$ & $\frac{pi}{2}$ & rad\\
    RAAN & $0$ & $2\pi$ & rad \\
    Arg. of Lat. & $0$ & $2\pi$ & rad\\
    True Anomaly & $0$ & $2\pi$& rad \\
    Desired Latitude & $-0.99i\times\frac{180}{\pi}$ & $0.99i\times\frac{180}{\pi}$ & $^\circ$\\
    Desired Longitude & $-180$ & $180$ & $^\circ$\\
    \bottomrule
\end{tabular}
\caption{Range of initial orbits, epochs and desired locations at entry interface for the guidance trajectories in the Monte Carlo simulation. }
\label{table:traj_param}
\end{table}

After the guidance trajectory is generated, it is then tracked by the spacecraft. The initial radial and in-track position of the spacecraft at the initial time is perturbed by adding a dispersion to the initial semi-major axis and true anomaly of the initial orbit of the tracking spacecraft corresponding to dispersions up to approximately $100$~m in the radial direction and $1$~km in the in-track direction. The density multiplier, $c_\text{drag}$, is sampled uniformly from the interval ranging from $0.75$ to $1.25$. The position and velocity of the spacecraft relative to the guidance are corrupted by noise at the beginning of each time step according to~\eqref{eq:meas} to produce synthetic GPS measurements that are used for the measurement update in the EKF. The noise is zero-mean and normally distributed according to the parameters in Table \ref{table:EKF_param} which are chosen to be representative of GPS measurement errors~\cite{Omar2019,Kovar2017}. 

\subsection{Algorithm Parameters}
Both the EKF and MPC are run with a sample time of $\Delta T \approx 600$ s which corresponds to a new state estimate and control input being computed and implemented approximately every 10 minutes or 9 times per orbit. The exact sample time is obtained by dividing the duration of the guidance trajectory by the desired sample time to obtain the number of time steps. This number is rounded to the nearest integer to compute the actual sample time. 

The EKF is run using the parameters in Table~\ref{table:EKF_param}. The matrix $\mbf{V}$ is chosen to match the noise characteristics used in generating the synthetic measurements while the process noise $\mbf{W}$ is treated as a tuning parameter and is selected to provide the desired level of performance in estimating the density error.

The MPC uses the parameters listed in Table~\ref{table:MPC_param}. The majority of these parameters are controller design parameters that are tuned to give the desired level of tracking performance. The control bounds, $C_{b,\text{min}}$ and $C_{b,\text{max}}$, are physical design parameters that depend on the drag control device being used. The values used here are selected to be representative of the range of ballistic coefficients that are achievable using a device similar to the ExoBrake \cite{Murbach2010}.

\begin{table}[!t]
\centering
\begin{tabular}{ccc}
    \toprule
    Parameter & Value &Units\\
    \midrule
    $\Delta T$ & $\approx$600 & s\\
    $\sigma_\text{pos}$ & 5 & m\\
    $\sigma_\text{vel}$ & 0.05 & m/s\\
    $\mbf{V}$ & $\text{diag}\left(\sigma_\text{pos}^2,\sigma_\text{pos}^2,\sigma_\text{vel}^2,\sigma_\text{vel}^2 \right )$& \\
    $\mbf{W}$ & $\text{diag}\left(1\times 10^{-5},1\times 10^{-6},2\times 10^{-5} \right )$& \\
    \bottomrule
\end{tabular}
\caption{Estimation filter parameters for the Monte Carlo simulations.}
\label{table:EKF_param}
\end{table}

\begin{table}[!t]
\centering
\begin{tabular}{ccc}
    \toprule
    Parameter & Value &Units\\
    \midrule
    $\Delta T$ & $\approx$600 & s\\
    $N$ & 36 & Time steps\\
    $N_{ss}$ & 5 & Substeps\\
    $\mbf{Q}_c$ & $\text{diag}\left(10, 1, 0, 0\right)$ & \\
    ${R_c}$ & $10^{10}$& \\
    $\mbf{Q}_{c,N}$ &$\mbf{P}_\text{Ric}$ (except for last $N$ time steps of reentry) & \\
    $\mbf{Q}_{c,N}$ &$\mbf{Q}_c$ (during last $N$ time steps of reentry) & \\
    $C_{b,\text{min}}$ & $0.025$ & m$^2$/kg\\
    $C_{b,\text{max}}$ & $0.1$ & m$^2$/kg\\
    \bottomrule
\end{tabular}
\caption{Controller parameters for the Monte Carlo simulations.}
\label{table:MPC_param}
\end{table}

Throughout the majority of the trajectory, the prediction horizon consists of $36$ time steps which corresponds to predicting approximately four orbits into the future. The terminal cost $\mbf{Q}_{c,N}$ is computed by solving the discrete-time algebraic Riccati equation (DARE), given by 
\beq    
    \label{eq:DARE}
    \mbf{A}^\trans \mbf{P}_\text{Ric}\mbf{A} -\left(\mbf{A}^\trans \mbf{P}_\text{Ric} \mbf{B}\right)\left( {R}_c+\mbf{B}^\trans\mbf{P}_\text{Ric}\mbf{B} \right)^{-1}\left(\mbf{B}^\trans \mbf{P}_\text{Ric}\mbf{A} \right)+\mbf{Q}_c-\mbf{P}_\text{Ric}= \mbf{0},
\eeq
where the solution $\mbf{P}_\text{Ric}$ is then used to define the terminal cost $\mbf{Q}_{c,N} = \mbf{P}_\text{Ric}$. For a linear time-invariant system with no constraints on the states or input, the use of the DARE solution perfectly reproduces the infinite horizon behavior of a linear quadratic regulator. The DARE solution is used here with the LTV dynamics and constrained input to approximate that behavior. The dynamics used in the DARE are the linearization at the beginning of the prediction horizon, $\mbf{A} = \mbf{A}_0$ and $\mbf{B} = \mbf{B}_0$. 

Later in the trajectory, when the prediction horizon reaches the final time of the guidance trajectory, the prediction horizon is reduced so that the end of the prediction horizon aligns with the end of the guidance trajectory. When the prediction horizon reaches the end of the guidance trajectory, the terminal cost is changed from that obtained from the solution to the DARE to a prescribed terminal cost $\mbf{Q}_{c,N} = \mbf{Q}_c$. This is done in order to prioritize a small tracking error at the final time of the guidance and produce a small error at the entry interface altitude. 

\subsection{Results}
\label{sec:Results}

\begin{figure}[t!]
	\centering
        \subfloat[]{
		\includegraphics[width=0.48\linewidth]{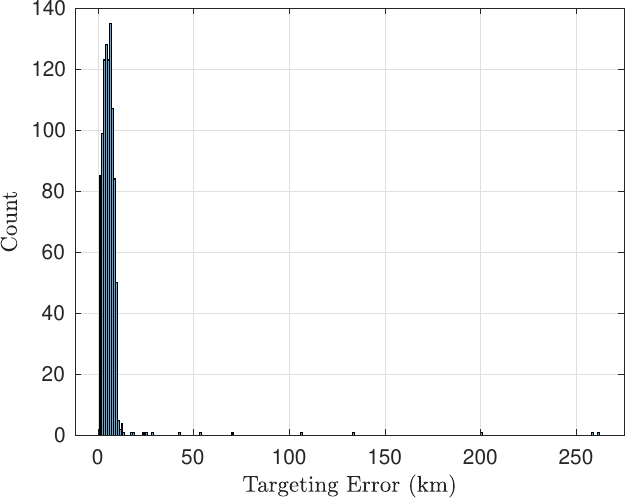} 

  }
        \subfloat[]{
        \includegraphics[width=0.48\linewidth]{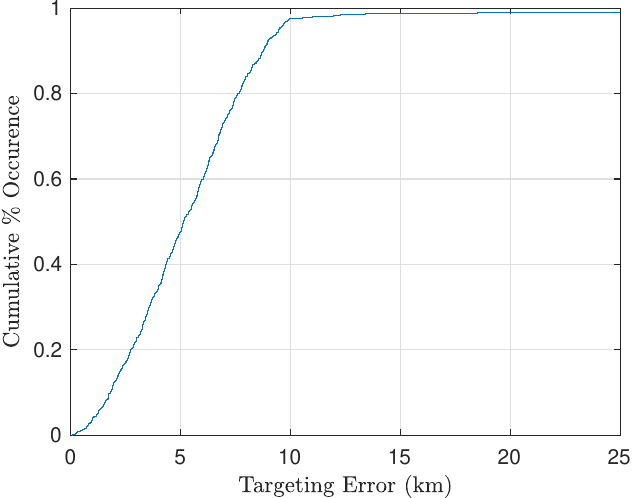} 
        }
	\caption{Monte Carlo results of the final guidance position errors as (a) a histogram and (b) a cumulative distribution function zoomed in on the errors below $25$~km.}
	\label{fig:targ_err}
\end{figure}

\begin{figure}[t!]
\centering
        \includegraphics[width=0.48\linewidth]{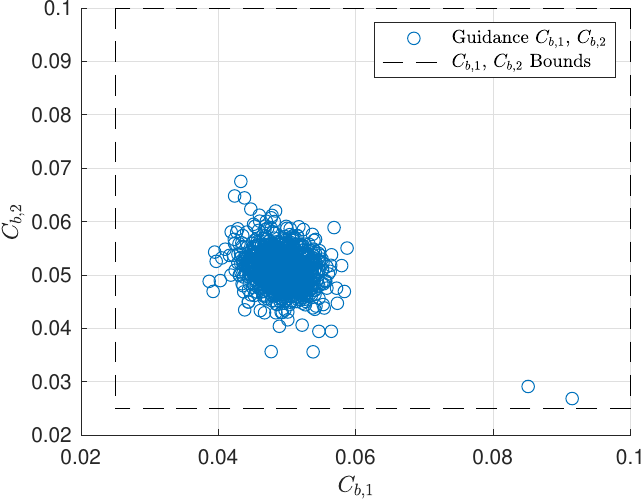} 
	\caption{Nominal ballistic coefficients for the Monte Carlo guidance trajectories (blue) and the spacecraft ballistic coefficient bounds (dashed black).}
	\label{fig:guidance_Cb}
\end{figure}

A guidance trajectory was successfully found for all 1000 cases. The desired final latitude, longitude and altitude for each case are used to obtain the desired final position of the spacecraft in the $ECEF$ frame. The targeting error of each guidance trajectory is then computed as the norm of the difference between desired final position of the spacecraft and final position of the guidance trajectory, both in the $ECEF$ frame. A histogram and cumulative distribution function of the targeting errors of each guidance trajectory are plotted in Fig.~\ref{fig:targ_err}. The mean targeting error is $6.31$ km and $97.7\%$ of cases have targeting error below the tolerance of $10$ km while $99.5\%$ of cases have targeting errors below $100$ km. The maximum targeting error is $262$ km. The ballistic coefficients that produce each guidance trajectory are shown in Fig.~\ref{fig:guidance_Cb}.  While the guidance generation algorithm is not a contribution of this work, the simplified implementation of the work by Omar successfully produced a guidance trajectory in all cases with performance comparable to the original implementation~\cite{Omar2017}. The vast majority of the targeting errors were small enough to safely target reentry away from populated areas, as shown in Fig.~\ref{fig:targ_err}. Additionally, our implementation is able to select trajectories corresponding to nominal ballistic coefficients that preserve more control authority to decrease the ballistic coefficient, as shown in Fig.~\ref{fig:guidance_Cb} compared to Fig.~\ref{fig:Cb_max_ctrbl}. While there are some outlier trajectories with large targeting errors, this can likely be improved using methods present in Omar's more sophisticated implementations, such as drag-work enforcement, back stepping and using a shrinking horizon. However, our simplified implementation was sufficient for the purpose of providing trajectories on which to test the proposed estimation and control methods. 

The guidance trajectories are tracked down to the desired entry interface altitude of $100$ km using the proposed estimation and control framework. If the spacecraft reaches the final altitude before the end of the guidance trajectory, then the simulation is terminated early when the spacecraft is at an altitude of $100$ km. Should the spacecraft be above the desired altitude when the guidance trajectory ends, the simulation continues until the spacecraft reaches the desired final altitude. In this case, the final tracking error at entry interface is then computed based on the final position of the guidance and the final position of the spacecraft at the entry interface altitude. Computing the final tracking error in this way is more indicative of the reentry accuracy compared to looking at the radial and in-track errors at the final time of the guidance trajectory. This is because tracking error in the radial and in-track directions at the final time of the guidance trajectory contribute differently to the reentry accuracy. Errors in the in-track direction at the final time of the guidance produce errors of similar magnitude throughout the trajectory after entry interface, whereas errors in the radial direction are magnified. By computing errors at the entry interface altitude instead of at the final time of the guidance, the relative contribution of in-track and radial errors at the final time on the entry accuracy is captured. 

\begin{figure}[t!]
	\centering
	\subfloat[]{
		\includegraphics[width=0.48\linewidth]{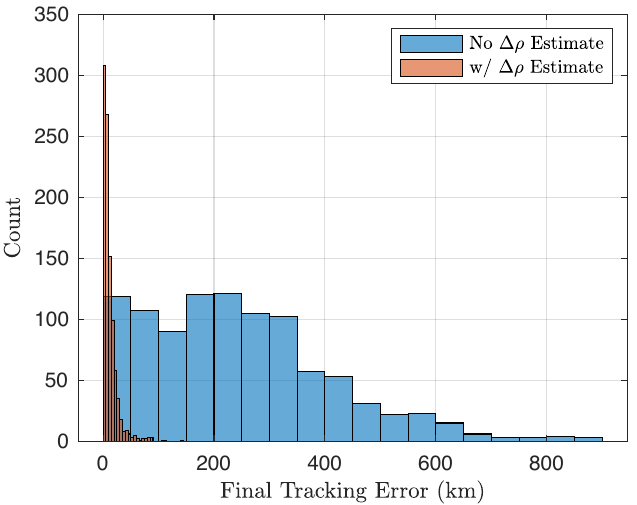}}
	\subfloat[]{
		\includegraphics[width=0.48\linewidth]{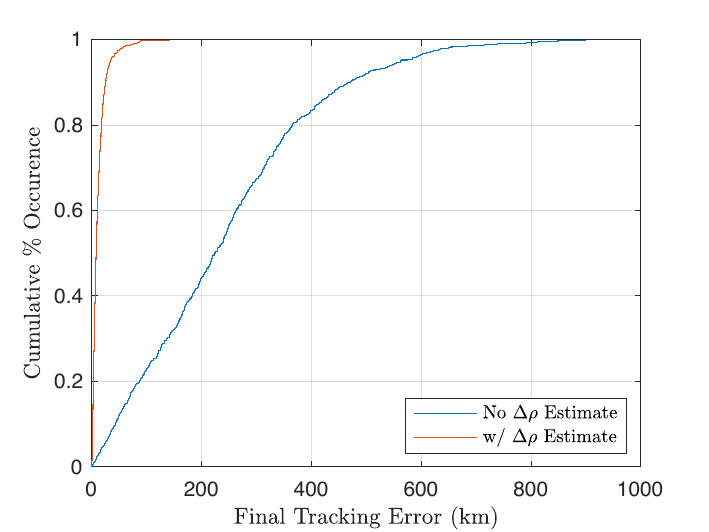}}
	\caption{Entry interface tracking errors with (red) and without (blue) estimated $\Delta\rho$ information in the control as (a) a histogram and (b) a cumulative distribution function.}
	\label{fig:track_err}
\end{figure}

Of the $1000$ simulations, 984 trajectories were successfully tracked down to entry interface in both the case where $\Delta\rho$ estimates were and were not used in the MPC controller, leading to a success rate of $98.4\%$. In the other $16$ cases, the tracking error exceeded the $100$ km threshold at some point in time and the simulations were terminated early. The final tracking errors at entry interface are shown in Fig.~\ref{fig:track_err}. Without including the estimates of $\Delta\rho$ in the control, the mean tracking error was $244$ km with $23.0\%$ of cases falling below 100 km. Including estimates of $\Delta\rho$ results in a mean tracking error of $12.1$ km, where $99.7\%$ cases fall below 100 km.  The combined MPC and EKF approach was able to guide the spacecraft accurately along the guidance trajectory and to the desired entry interface location in almost every case. Of the cases that were terminated early, several coincided with periods of volatile atmospheric conditions associated with solar storms and flares such as those in 2005. Under such extreme solar conditions, the atmospheric density changes to an extent that the drag control device inevitably becomes saturated and the guidance trajectory is not trackable. The tracking errors for the successful cases, as shown in Fig.~\ref{fig:track_err}, were sufficiently small for the spacecraft to safely target a reentry over the ocean with a success rate of $98.4\%$. This demonstrates the capability of the proposed targeted reentry control method for the deorbit of small satellites with mission requirements of safely splashing down in the ocean, such as the HyCUBE concept~\cite{anderson2021preliminary,gardi2024simulation}.

\begin{figure}[t!]
	\centering
		\includegraphics[width=0.48\linewidth]{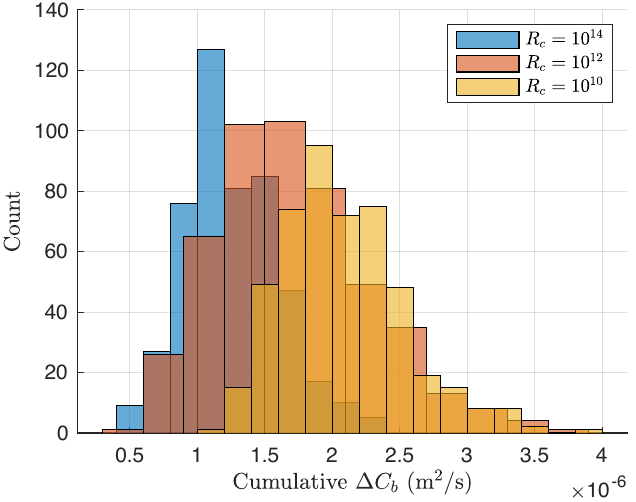}
	\caption{Actuator use per unit time with different tunings of the MPC controller.}
	 \label{fig:tune_act}
\end{figure}

\begin{figure}[t!]
	\centering
        \subfloat[]{
        \includegraphics[width=0.48\linewidth]{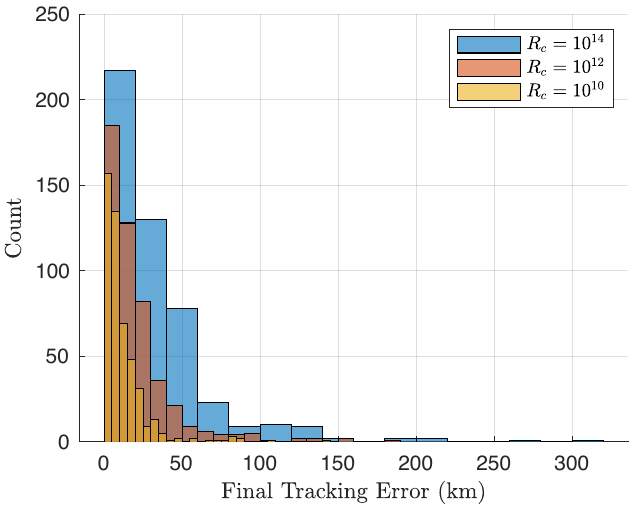}
        }
         \subfloat[]{
        \includegraphics[width=0.48\linewidth]{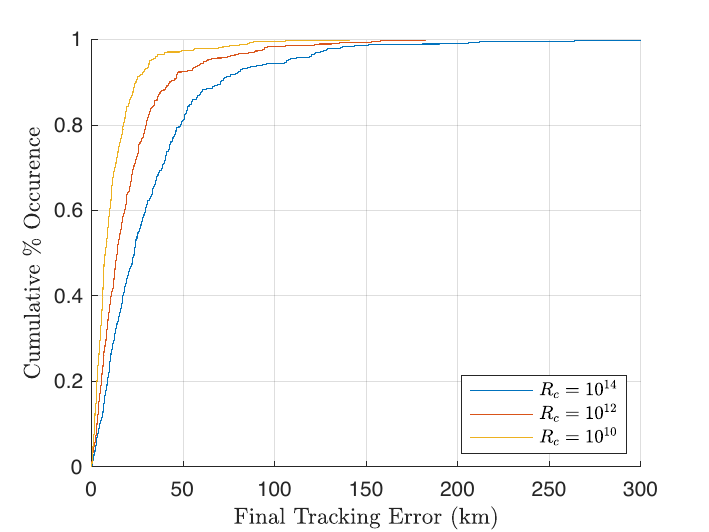}
        }
	\caption{Entry interface tracking errors with different tunings of the MPC controller as (a) a histogram and (b) a cumulative distribution function.}
	\label{fig:tune_err}
\end{figure}

Monte Carlo simulations with the estimates of $\Delta\rho$ are repeated for three different values of $R_c$ in the MPC controller to observe the effect of tuning on the control effort and final tracking error. Only the first $500$ randomly-generated trajectories are considered and the gravity model is set to the same fidelity used during the guidance generation to decrease run time. To obtain a sense of how much the drag device is used, the sum of all of the $\Delta C_b$ for a particular trajectory is divided by the duration of the corresponding trajectory. Fig.~\ref{fig:tune_act} shows the actuator use per unit time for all of the trajectories. Fig.~\ref{fig:tune_err} shows a histogram and cumulative distribution function of the final tracking errors for the three different tunings.  Increasing $R_c$ to increase the cost of changing the control input reduces the total change in the ballistic coefficient throughout the trajectories, as can be seen in Fig.~\ref{fig:tune_act}. However, the reduced actuation results in larger tracking errors when the spacecraft reaches entry interface, as evident in Fig.~\ref{fig:tune_err}. This demonstrates that the controller is able to be adjusted to give the desired balance of tracking performance and actuator use.

\begin{figure}[t!]
	\centering
		\includegraphics[width=0.6\linewidth]{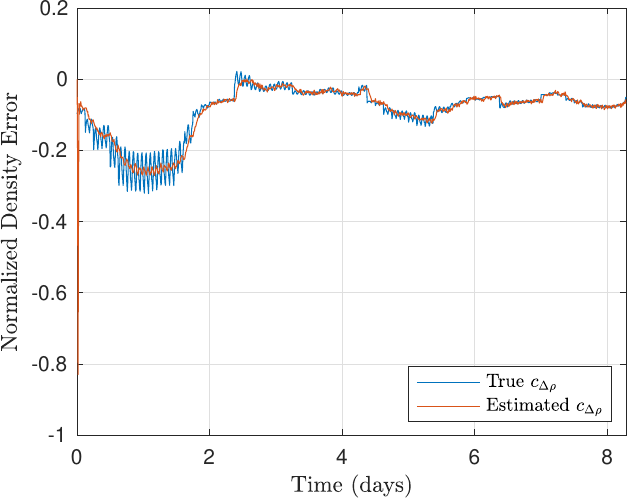}
	\caption{True value of the normalized density prediction error (blue) and the estimated value (red) from a single trial of the Monte Carlo simulation.}
	\label{fig:c_drho}
\end{figure}

\begin{figure}[t!]
	\centering
        \subfloat[]{
        \includegraphics[width=0.48\linewidth]{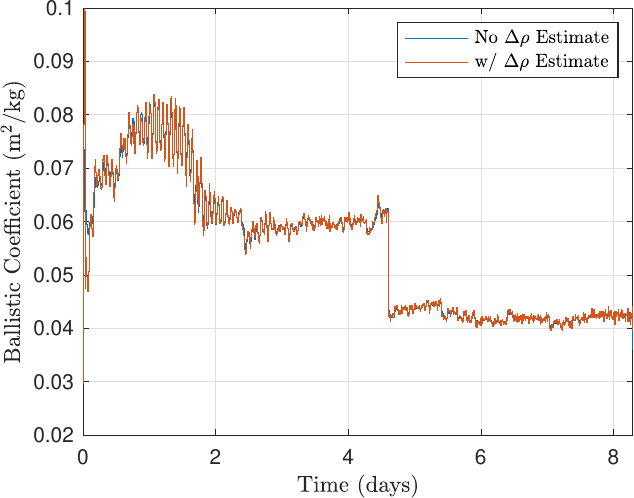}
        \label{fig:Cb_cmd_mc}
        }
        \subfloat[]{
        \includegraphics[width=0.48\linewidth]{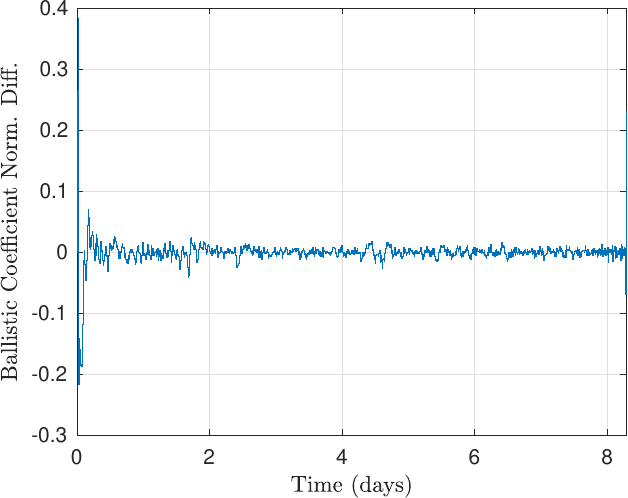}
        \label{fig:Cb_cmd_mc_diff}
        }
	\caption{(a) Ballistic coefficient commanded by MPC without (blue) and with (red) estimated density information and (b) the normalized difference between them.}
	\label{fig:Cb_cmd_mc_tot}
\end{figure}

\begin{figure}[t!]
	\centering
		\includegraphics[width=0.6\linewidth]{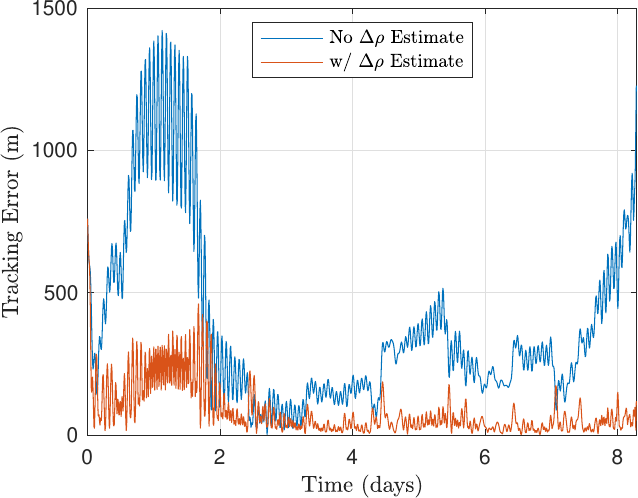} 
	\caption{Tracking error for MPC without (blue) and with (red) using the estimated density error information.}
	\label{fig:drho_v_nodrho}
\end{figure}

The true value of $c_{\Delta\rho}$ for one particular case from the Monte Carlo is shown in Fig.~\ref{fig:c_drho} along with the value estimated by the EKF. The control inputs produced by MPC in both the case where the estimated density information is not used and the case where it is used are shown in Fig.~\ref{fig:Cb_cmd_mc}. The tracking errors produced in these two cases are shown in Fig.~\ref{fig:drho_v_nodrho}. As shown in Fig.~\ref{fig:track_err}, the inclusion of the estimated density error from the EKF drastically improves the tracking performance of the MPC controller. The single case in Fig.~\ref{fig:c_drho} demonstrates that the EKF is able to estimate an average value of the normalized density prediction error, although it does not capture the oscillations that occur with the same period of the spacecraft orbit. This estimated knowledge of the atmosphere is able to remove a significant amount of the tracking error that builds up with out this knowledge, as shown in Fig.~\ref{fig:drho_v_nodrho}. This, however, requires the ability to adjust the ballistic coefficient of the spacecraft very precisely. As shown in Fig.~\ref{fig:Cb_cmd_mc}, the commanded ballistic coefficients are only very slightly different than those commanded without the density error estimate. The difference between these ballistic coefficients normalized by the ballistic coefficient without estimated knowledge of the density error is included in Fig.~\ref{fig:Cb_cmd_mc_diff}.

\section{Conclusions and Future Work}
\label{sec:Conclusion}
A model predictive controller that leverages estimates of the atmospheric density was further developed to improve its tracking performance. This control scheme was combined with a filter that is capable of estimating the density of the atmosphere compared to an a priori prediction using measurements of the motion of the spacecraft in order to reduce the tracking errors caused by errors in predicting atmospheric density. A Monte Carlo simulation campaign was conducted to evaluate the performance of the combined estimation and control framework over a wide range of low-Earth orbit conditions and subject to realistic uncertainties in the gravitational force and atmospheric density. The simulation campaign showed that the estimation and control framework is able to reliably guide the spacecraft to a targeted reentry accurately enough to avoid posing a hazard to people and property on the ground. 

Opportunities exist to further improve the performance of the estimation and control framework. More accurate estimates of the atmospheric density would improve the tracking performance of the MPC by more accurately rejecting the disturbance due to atmospheric density prediction error. The current estimation and control framework operates the filter at the same rate as the control, however, the filter may be run at an increased rate to obtain more accurate state and density estimates. Additionally, an improved process model for how the density prediction error evolves that captures the oscillations on the period of the orbit would similarly increase the accurate of the estimates. The tracking performance can also be improved through changes to the MPC algorithm. Tracking performance can be improved while reducing actuator usage by varying the control frequency throughout the trajectory. When the spacecraft is at a high altitude, the actuation frequency can be reduced to reduce wear on the actuator. At low altitudes when the spacecraft is near reentry, the actuation frequency can be increased to reduce tracking error. Tracking performance can also be improved by using a more accurate method of discretizing the dynamics, such as a first-order hold instead of the zero-order hold. 

\section*{Appendix}

This appendix provides a detailed summary of the simplified guidance trajectory generation method based on work by Omar~\cite{Omar2017conf,Omar2019}. The guidance scheme assumes that the ballistic coefficient of the spacecraft is nominally modulated a single time, as opposed to multiple changes in ballistic coefficient or continuous modulation. In accordance with this assumption, the spacecraft has an initial ballistic coefficient, $C_{b1}$, which is held constant until time $t_{\text{swap}}$ at which point the ballistic coefficient of the spacecraft changes to $C_{b2}$. 
The guidance problem is then a matter of finding the values of the guidance parameters $C_{b1}$, $C_{b2}$ and $t_{\text{swap}}$ that result in proper entry interface targeting.

The guidance algorithm makes use of analytical approximations that describe how a given, numerically propagated trajectory is changed by a change in the guidance parameters. This enables the trajectory to be iteratively refined beginning with an initial guess of the guidance parameters. In each iteration, the previous values of the guidance parameters are used to numerically propagate the trajectory, which can then be used to compute improved values of the guidance parameters that produce a trajectory with an entry interface location that is closer to the desired location. %

Consider a spacecraft with a ballistic coefficient of $C_{b10}$ in an orbit that decays from an initial to a final value of semi-major axis. During this decay, the spacecraft has a total change in true anomaly of $\Delta\theta_{10}$ over a change in time of $\Delta t_{10}$. These values are obtained by simulating the trajectory with a ballistic coefficient of $C_{b10}$. Omar's work shows that under the approximations of a circular orbit and an exponential atmospheric density profile, if the ballistic coefficient changes from $C_{b10}$ to $C_{b1}$, then the change in true anomaly and change in time during the decay between the same
initial and final semi-major axes will change to approximately $\Delta\theta_1$ and $\Delta t_1$ according to
\beq
    \label{eq:basic_targ_theta}
    \Delta\theta_1 = \Delta\theta_{10}\frac{C_{b10}}{C_{b1}}
\eeq
and 
\beq
    \label{eq:basic_targ_t}
    \Delta t_1 = \Delta t_{10}\frac{C_{b10}}{C_{b1}}.
\eeq
These analytical approximations are applied to different phases of the deorbit to derive relationships between the guidance parameters and the location of the spacecraft at the entry interface altitude given a previous numerically-propagated trajectory from the previous iteration or initialization.

The algorithm is initialized with an initial guess of the guidance parameters and the resulting trajectory. Then the analytical approximations are used to compute new estimates of the guidance parameters that change the previously propagated trajectory in order to correct errors in the targeting in distinct latitude and longitude targeting steps.

\subsubsection{Algorithm Initialization}
The guidance generation begins by simulating the orbital decay of the spacecraft with a constant ballistic coefficient of $C_{b,\text{max}}$ resulting in the time of flight for this trajectory $t_{f,0}$. This provides a feasible range for $t_\text{swap}$ that must fall between $t = 0$ and $t = t_{f,0}$ for a spacecraft with $C_{b1} = C_{b,\text{max}}$. The initial guess of the guidance parameters is then made as $C_{b1} = C_{b,\text{max}}$, $C_{b2} = C_{b,\text{min}}$, and $t_\text{swap} = \frac{t_{f,0}}{2}$. These guidance parameters are then numerically propagated to obtain a trajectory which serves as the basis of the first iteration, which computes a change in $t_\text{swap}$ in order to reduce the latitude error at entry interface. 

\subsubsection{Latitude Targeting}
An angle $\phi$ is defined in the orbital plane from the ascending node of the orbit to the position of the spacecraft. There are two angles, $\phi_{d1}$ and $\phi_{d2}$, that correspond to any desired final latitude. The trajectory from the previous iteration will sweep out a total angle of $\Delta\phi_t$. The trajectory can be altered to land at the desired latitude by changing $\Delta\phi_t$ by a $\Delta\phi_d$ that is chosen to ensure that the spacecraft is at an angle of $\phi_{d1}$ or $\phi_{d2}$ at the final time. This results in a desired total swept angle of $\Delta \phi_{t,d} = \Delta \phi_t + \Delta \phi_d$. It is shown by Omar \cite{Omar2017conf} that this change in total angle can be obtained by changing when the switch between ballistic coefficients occurs by 
\beq
    \Delta t_\text{swap} = \Delta\phi_d\frac{C_{b2}}{\omega_{2,avg}\left( C_{b2}-C_{b1} \right)},
\eeq
where $\omega_{2,avg}$ is an average angular velocity calculated from the trajectory from the previous iteration. A change in $\Delta\phi_{t,d}$ of an integer multiple of $2\pi$ radians results in the spacecraft reaching the same latitude, but a integer number of orbits earlier or later. As a result, if the spacecraft starts at a sufficiently high altitude and has sufficient control authority, there are many possible values of $\Delta\phi_{t,d}$ that target the desired latitude leading to many feasible values of $t_\text{swap}$. If the initial altitude is too low or the control authority is too small, there may be no feasible choices of $t_\text{swap}$.

\subsubsection{Longitude Targeting}
After latitude targeting, the deorbit trajectory will sweep out an angle, $\Delta\phi_{t,d}$, that ensures the desired latitude is reached. However, because the latitude targeting step does not address longitude, there will be a residual longitude error that requires reduction. The longitude error, $e_\text{long}$ is estimated by using the value of $t_\text{swap}$ from latitude targeting along with~\eqref{eq:basic_targ_theta} and~\eqref{eq:basic_targ_t} to estimate the new orbital lifetime. The new final longitude after latitude targeting can then be predicted based on the final longitude of the trajectory from the previous iteration, the predicted change in orbital lifetime due to $\Delta t_\text{swap}$, the rotation rate of the Earth and the precession rate of the orbit. The longitude error is then computed based on this predicted final longitude. This longitude error is addressed by changing the orbital lifetime while preserving the total angle swept out during deorbit. This will ensure that the spacecraft lands at the same latitude, but at a different time. Because the spacecraft reaches entry interface at a different time, the Earth will be in a different orientation below the orbit of the spacecraft, which itself will have precessed by a different amount. Changing the orbital lifetime, $\Delta t_t$, by
\beq
    \Delta t_d = \frac{e_\text{long}}{\omega_e - \omega_\text{RAAN}}, 
\eeq
where $\omega_\text{RAAN}$ is the average rate of change of the right ascension of the ascending node throughout the trajectory due to the J2 perturbation, results in a new trajectory with a reduced longitude error. This new trajectory has an orbital lifetime of $\Delta t_{t,d} = \Delta t_t + \Delta t_d$.

The change in orbital lifetime of $\Delta t_d$ is accomplished without changing $\Delta\phi_t$ through a change in $C_{b1}$ and $C_{b2}$~\cite{Omar2019}. Given a numerically propagated trajectory from the previous iteration with an initial ballistic coefficient of $C_{b10}$ that changes to $C_{b20}$ at $t_\text{swap}$, new ballistic coefficients are estimated that give a trajectory with $\Delta\phi_{t,d}$ and $\Delta t_{t,d}$ through
\beq
    \label{eq:Cb_2_long}
    C_{b2} = C_{b20}\frac{\Delta t_{20}\Delta\theta_{10} - \Delta t_{10}\Delta\theta_{20}}
    {\Delta t_{t,d}\Delta\theta_{10}\Delta t_{10}\Delta\theta_{t,d}},
\eeq
and
\beq
    \label{eq:Cb_1_long}
    C_{b1} = C_{b10} \frac{\Delta\theta_{10}C_{b2}}{\Delta\theta_{t,d}C_{b2} -\Delta\theta_{20}C_{b20}},
\eeq
where $\Delta\theta_{10}$ and $\Delta\theta_{20}$ are the total amount of true anomaly swept through from $t = 0$ to $t = t_\text{swap}$ and $t_\text{swap}$ to the final time of the numerically propagated trajectory, respectively. Similarly, $\Delta t_{10}$ and $\Delta t_{20}$ are the duration of the trajectory phases before and after $t_\text{swap}$ in the propagated trajectory from the previous iteration. During the latitude targeting step, a value of $t_\text{swap}$ must be chosen such that the computation of $C_{b1}$ and $C_{b2}$ results in values ranging between $C_{b,\text{min}}$ and $C_{b,\text{max}}$. If no such value of $t_\text{swap}$ exists, then the spacecraft does not have sufficient control authority to target the desired landing site from the initial altitude.

These equations assume that the change between ballistic coefficients occurs at the same semi-major axis. However, because $C_{b1}$ has changed, the spacecraft will reach this semi-major axis at a different time. Therefore $t_\text{swap}$ must be changed to accommodate this change in $C_{b1}$. The new swap time is estimated as 
\beq
    t_\text{swap,new} = t_\text{swap,old}\frac{C_{b10}}{C_{b1}}.
\eeq
These new values of the guidance parameters can be used to numerically propagate the trajectory of the spacecraft from the initial condition to obtain a new guess of the guidance trajectory, to serve as the basis of the next iteration. That is, $C_{b1}$ and $C_{b2}$ become $C_{b10}$ and $C_{b20}$ for the next iteration while $t_\text{swap,new}$ becomes $t_\text{swap}$.

\subsubsection{Iteration}
\label{sec:GuidanceIteration}

There are many choices of $t_\text{swap}$ that can be used to perform latitude targeting in the final step of the iterative guidance algorithm. The value of $t_\text{swap}$ can be chosen to minimize the residual longitude error~\cite{Omar2017conf}, provide the largest amount of controllability margin about the desired longitude~\cite{Omar2017}, or remain as centered as possible within the range of feasible ballistic coefficients (a contribution of this work described in Section~\ref{sec:GuidanceGeneration}). Once $t_\text{swap}$ is chosen via one of these methods, $t_\text{swap}$ and the corresponding $C_{b1}$ and $C_{b2}$ are then numerically simulated, producing a trajectory that is used as the basis of the next iteration. This process repeats until a trajectory is produced with a final targeting error of less than some tolerance or until a maximum number of iterations is reached. The position and velocity of the spacecraft along this trajectory, $\mbf{r}^{gp}_{ECI}(t)$ and $\mbf{v}^{gp/ECI}_{ECI}(t)$, serve as the guidance trajectory for the estimation and control algorithms to track while $C_{b1}$, $C_{b2}$ and $t_\text{swap}$ form the nominal control strategy where the guidance ballistic coefficient is
\beq
    C_{b,g}(t) = \begin{cases} 
      C_{b1} & 0\leq t \leq t_\text{swap} \\
      C_{b2} & t_\text{swap} < t\leq t_f
   \end{cases}
\eeq
While the way in which $t_\text{swap}$ is selected differs from previous guidance generation work in order to preserve more control authority for feedback control, guidance generation is not considered a major contribution of this work. Guidance generation is performed in order to produce trajectories for the purpose of evaluating the estimation and control algorithms. As such, this implementation is simplified compared to previous work~\cite{Omar2017conf,Omar2019}, which include more sophisticated methods that increase the performance and robustness of such guidance generation algorithms. 

\section*{Funding Sources}

This research was supported in part by the Air Force Office of Scientific Research (AFOSR), United States under Grant No. FA9550-19-1-0308. The views and conclusions contained herein are those of the authors and should not be interpreted as necessarily representing the official policies or endorsements, either expressed or implied, of the AFOSR or the U.S. Government.  This work was also supported by an award from MnDRIVE and the Minnesota Informatics Institute.

\section*{Acknowledgments}

The authors would like to thank Dr. Demoz Gebre-Egziabher for his insight into the Kalman filter used to perform the atmospheric density estimation.

\bibliography{Bibliography_edit}

\end{document}